\documentclass{jpp}

\usepackage{graphicx}
\usepackage{natbib}

\usepackage{hyperref} 
 \usepackage{enumerate}  

\usepackage{stmaryrd} 
\usepackage{mathcomp} 

\usepackage{bm} 

\ifCUPmtlplainloaded \else
  \checkfont{eurm10}
  \iffontfound
    \IfFileExists{upmath.sty}
      {\typeout{^^JFound AMS Euler Roman fonts on the system,
                   using the 'upmath' package.^^J}%
       \usepackage{upmath}}
      {\typeout{^^JFound AMS Euler Roman fonts on the system, but you
                   dont seem to have the}%
       \typeout{'upmath' package installed. JPP.cls can take advantage
                 of these fonts, if you use 'upmath' package.^^J}%
      }
  \else
  \fi
\fi

\ifCUPmtlplainloaded \else
  \checkfont{msam10}
  \iffontfound
    \IfFileExists{amssymb.sty}
      {\typeout{^^JFound AMS Symbol fonts on the system, using the
                'amssymb' package.^^J}%
       \usepackage{amssymb}%
         \let\leq=\leqslant
         
      }{}
  \fi
\fi

\ifCUPmtlplainloaded \else
  \IfFileExists{amsbsy.sty}
    {\typeout{^^JFound the 'amsbsy' package on the system, using it.^^J}%
     \usepackage{amsbsy}}
    {\providecommand\boldsymbol[1]{\mbox{\boldmath $##1$}}}
\fi

\providecommand\bnabla{\boldsymbol{\nabla}}
\providecommand\bcdot{\boldsymbol{\cdot}}

\newsavebox{\astrutbox}
\sbox{\astrutbox}{\rule[-5pt]{0pt}{20pt}}

\providecommand\grad{\bnabla} 
\providecommand\divv{\bnabla\bcdot} 
\providecommand\cross{\!\boldsymbol{\times}\!} 
\providecommand\curl{\bnabla\cross} 
\providecommand\dotv{\bcdot} 

\renewcommand{\d}{\mathrm{d}}
\newcommand{\esup}[1]{\boldsymbol{e}^{#1}} 

\newcommand{\const}{{\mathrm{const}}}
\newcommand{\eqref}[1]{eq.~$\!$(\ref{#1})} 
\newcommand{\comp}{\,{\scriptstyle{\circ}}\,}

\renewcommand{\vec}[1]{\bm{#1}} 
\newcommand{\vrm}[1]{{\boldsymbol{#1}}} 
\newcommand{\vsf}[1]{\mathsfbi{#1}} 

\newcommand{\jump}[1]{\left\llbracket  #1 \right\rrbracket} 

\newcommand{\muSI}{\tcmu_0}

 \title[Variational formulation of RxMHD and MRxMHD]{Variational formulation of relaxed and  multi-region relaxed magnetohydrodynamics}

\author[R.~L. Dewar, Z. Yoshida, A. Bhattacharjee and S.~R. Hudson]%
{R.\ns L.\ns D\ls E\ls W\ls A\ls R$^1$%
 \thanks{Email address for correspondence: robert.dewar@anu.edu.au.},\ns%
Z.\ns Y\ls O\ls S\ls H\ls I\ls D\ls A$^2$\break
A.\ns B\ls H\ls A\ls T\ls T\ls A\ls C\ls H\ls A\ls R\ls \ls J\ls E\ls E$^3$ \and S.\ns R.\ns  H\ls U\ls D\ls S\ls O\ls N$^3$
 }
\affiliation{$^1$Centre for Plasmas and Fluids, Research School of Physics \& Engineering, The Australian National University, Canberra, ACT 2601, Australia\\[\affilskip] 
$^2$Graduate School of Frontier Sciences, University of Tokyo, Kashiwa, Chiba 277-8561, Japan\\[\affilskip]
$^3$Princeton Plasma Physics Laboratory, PO Box 451, Princeton NJ 08543, USA}

\date{v2 \today} 

\begin{document}

\maketitle
\begin{abstract}
Ideal magnetohydrodynamics (IMHD) is strongly constrained by an infinite number of microscopic constraints expressing mass, entropy and magnetic flux conservation in each infinitesimal fluid element, the latter preventing magnetic reconnection. By contrast, in the Taylor relaxation model for formation of macroscopically self-organized plasma equilibrium states, all these constraints are relaxed save for global magnetic fluxes and helicity. A Lagrangian variational principle is presented that leads to a new, fully dynamical, \emph{relaxed magnetohydrodynamics} (RxMHD), such that all static solutions are Taylor states but also allows flow. By postulating that some long-lived macroscopic current sheets can act as barriers to relaxation, separating the plasma into multiple relaxation regions, a further generalization, \emph{multi-region relaxed magnetohydrodynamics} (MRxMHD) is developed.
\end{abstract}

\begin{PACS}
Authors should not enter PACS codes directly on the manuscript, as these must be chosen during the online submission process and will then be added during the typesetting process (see http://www.aip.org/pacs/ for the full list of PACS codes)
\end{PACS}

\section{Introduction}\label{sec:intro}
\label{sec:Intro}
	
The coarse-grained, fluid-like dynamical behaviour of highly conducting, magnetized plasmas in the laboratory, in stars such as the sun, and in space, can often be described by variants of \emph{magnetohydrodynamics} (MHD).

The one-fluid, non-dissipative model most commonly used, \emph{ideal magnetohydrodynamics} (IMHD), appears deceptively simple but is strongly constrained by an infinite number of microscopic constraints expressing the detailed conservation of mass, entropy and magnetic flux ``frozen'' into each infinitesimal fluid element.  Physically, these are very restrictive constraints, e.g. the ``entropy freezing'' constraint prevents dynamical temperature equilibration along magnetic field lines, and the ``flux-freezing'' constraint prevents changes in the topology of magnetic field lines, thus preventing magnetic reconnection, island formation, or formation of chaotic lines. These constraints also give rise to mathematical problems due to a tendency for singularities to form when systems are perturbed away from simple geometries with a continuous symmetry, \cite{Grad_67,Cary_Kotschenreuther_85,Hegna_Bhattacharjee_89,Bhattacharjee_etal_95, Hudson_etal_12b,Helander_14,Loizu_etal_15a,Loizu_etal_15b}. It is the aim of this paper to formulate an alternative self-consistent, non-dissipative single-fluid model for toroidal plasmas that is simpler than IMHD yet is less physically restrictive and better posed mathematically in general geometries. 

The most powerful and general way to formulate a non-dissipative field theory, see e.g. p. 53ff of \cite{Goldstein_80}, is to postulate a Lagrangian density $\mathcal{L}$ and to derive the dynamical equations for all fields from the action, $\mathcal{S} = \int\!\d t\!\int_\Omega\mathcal{L}\, \d^3x$, by appealing to Hamilton's Principle. That is, by requiring that its first variation, $\delta \mathcal{S}$, vanish for all variations of the independent fields in the system region $\Omega$. The equations for these fields are the resulting Euler--Lagrange equations, which, as they are all derived from the one scalar functional $\mathcal{S}$, are automatically self-consistent. Furthermore, conservation equations can be derived very generally by applying Noether's Theorem [see e.g. p. 555ff of \cite{Goldstein_80}, also \cite{Charidakos_etal_14} and references therein], based on the continuous symmetries of the system. \emph{Holonomic} constraints can be handled by expressing variations of dependent fields in terms of those of the independent fields, and \emph{non-holonomic} constraints by augmenting the Lagrangian density using Lagrange multipliers. Our modified MHD is based on the same Lagrangian as IMHD, but uses a much-reduced set of constraints, a small subset of those implicit in IMHD.

In IMHD and our modifications of it, on the boundary $\partial\Omega$ of the overall plasma region $\Omega$ 
(and on current sheets separating plasma subregions), 
the magnetic field $\vrm{B}$ is constrained everywhere to be a tangent vector
\begin{equation}\label{eq:tangential}
	\vrm{n}\dotv\vrm{B} = 0 \;,
\end{equation}
where $\vrm{n}$ is the unit normal at each point on $\partial\Omega$. Physically, this corresponds to the assumption of confinement within a perfectly conducting wall [but not necessarily a rigid wall if one wishes to model, for example, the response to an externally imposed perturbation by switching on boundary ripple,  \cite{Hahm_Kulsrud_85,Dewar_Bhattacharjee_Kulsrud_Wright_13,Comisso_Grasso_Waelbroeck_15a,Comisso_Grasso_Waelbroeck_15b}]. In this paper we assume for simplicity that the wall has no gaps, so that the wall completely shields the plasma from penetration of externally generated magnetic fluxes, but this restriction is not essential for a Lagrangian formulation to be possible, \cite{Dewar_78a,Hosking_Dewar_15}, and would need to be lifted if one wished to consider Ohmic current drive or helicity injection.

\begin{figure}
   \centering
		\includegraphics[width = 0.4\textwidth]{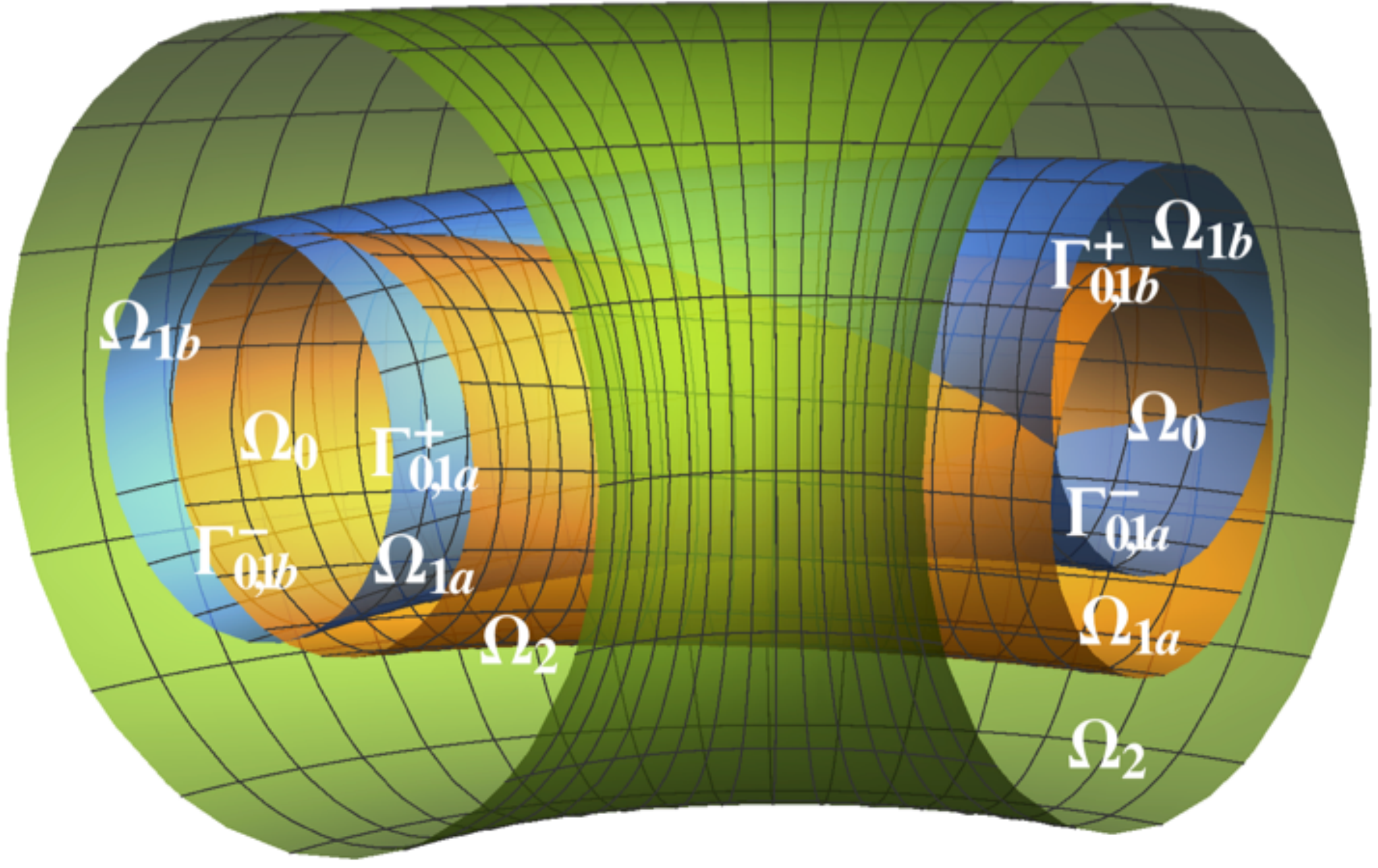} 
\caption{Some possible relaxation regions, discussed in Sec.~\ref{sec:Generalizations}. The innermost region,  $\Omega_0$, is a simple genus-1 toroid bounded by a torus $\partial\Omega_0$, formed from  $\Gamma_{0,1a}^-$ and $\Gamma_{0,1b}^- $  (blue and orange), the inward faces of the inner separatrix current sheets, \cite{Waelbroeck_89,Wang_Bhattacharjee_95}, of an $m=2,n=1$ magnetic island chain. The helical interiors of the two islands of the chain, $\Omega_{1a}$ and $\Omega_{1b}$, are also genus-1 toroids, bounded by the tori obtained by joining $\Gamma_{0,1a}^+$ with $\Gamma_{1a,2}^-$, and $\Gamma_{0,1b}^+$ with $\Gamma_{1b,2}^-$, where $\Gamma_{1a,2}^-$ and $\Gamma_{1b,2}^-$ are the inward faces of the outer separatrix current sheets (blue and orange).  (Colour online.)}
\label{fig:Topologies}
\end{figure}

In the following we will partition $\Omega$ into interacting subregions $\Omega_i$, on the boundaries of which the tangential-$\vrm{B}$  constraint \eqref{eq:tangential}
is also enforced, but tangential discontinuities due to current sheets on these interfaces are allowed. (A vacuum region $\Omega_{\rm v}$ in which no IMHD invariants other than total magnetic fluxes are assumed, can also be included between the wall and a plasma-vacuum interface.)

We also make the topological assumption that the plasma regions $\Omega_i$ are toroids, so their boundaries $\partial\Omega$ are tori,  
where by (generalized) \emph{tori} we mean 2-dimensional 
surfaces without boundary, conceivably multi-handled, and by \emph{toroids} we mean 3-dimensional volumes bounded by a single torus (which we term a \emph{simple toroid}), or by an inner and an outer torus (an \emph{annular toroid}). (See Fig.~\ref{fig:Topologies}.)

The single-fluid, non-dissipative MHD equations are encapsulated in the Lagrangian density, \cite{Newcomb_62,Dewar_70},
\begin{equation}\label{eq:L}
	\mathcal{L}^{\rm MHD} \equiv \frac{1}{2}\rho v^2- \frac{p}{\gamma-1} - \frac{B^2}{2\muSI} \;,
\end{equation}
where $\rho$ is the mass density, $p$ is the pressure, and $\muSI$ is the permeability of free space. The plasma is treated thermodynamically as an ideal gas with an isentropic (adiabatic) equation of state, $p/\rho^\gamma = \const$ (in IMHD this being applied microscopically in each fluid element). 

The ideal (IMHD) equations of motion follow, \cite{Newcomb_62,Dewar_70}, from \eqref{eq:L} by applying Hamilton's principle of stationary action, treating trial displacements of fluid elements from their physical positions as an arbitrarily variable vector field, to which the variations of density $\rho$, and pressure $p$ (or entropy density) and magnetic field ${\bf B}$, are holonomically constrained microscopically. A (non-canonical) Hamiltonian formulation of IMHD, in which constraints appear as degeneracies of Poisson brackets (some of which can be integrated as Casimir invariants), is also possible, \cite{Morrison_98,Yoshida_Dewar_12}, but the Lagrangian approach provides a more convenient starting point for finding a modified magnetohydrodynamics.  A Hamiltonian formulation could be derived from our new Lagrangian formulation, but this is not pursued in the present paper.

In strong contrast to IMHD, Taylor's relaxed equilibrium model, \cite{Taylor_74}, relaxes all the IMHD constraints save for conservation of toroidal magnetic flux and global magnetic helicity (which are IMHD invariants), leading to a very low-energy ``relaxed'' equilibrium state. Such a ``Taylor state'' is a special static solution of the IMHD equations, but is dynamically inaccessible via IMHD from \emph{arbitrary} initial states because of
IMHD's infinity of extra constraints. To elevate Taylor's  \emph{static} relaxed equilbrium theory to a relaxed magnetohydro-\emph{dynamics} (RxMHD), we use the Lagrangian \eqref{eq:L} and the same holonomic density constraint as for IMHD,  \cite{Newcomb_62,Dewar_70}, but treat the pressure and magnetic field as \emph{independently} variable fields subject only to conservation of global flux(es), entropy and magnetic helicity within $\Omega$, and the holonomic tangential-$\vrm{B}$  constraint \eqref{eq:tangential}. 
This makes the Taylor equilbrium state always dynamicallly accessible within RxMHD.

Physically, the Taylor model is designed to predict the final \emph{macroscopic}%
\footnote{By ``macroscopic'' we mean a coarse-grained description in which the small-scale reconnection processes are spatially and temporally unresolved, and also a description that is only correct in the limit that the non-dimensional conductivity parameter (Lundquist number) increases toward infinity, the dynamo flows being assumed to be higher order in inverse Lundquist number so that the Taylor state is a static, force-free MHD equilibrium.} 
self-organized state to which a highly conducting plasma will evolve, provided it has a stochastic mechanism for breaking the ``freezing in'' of magnetic field by the plasma, \cite{Rusbridge_91,Qin_Liu_Li_Squire_12}, and for generating magnetic field by small ``dynamo'' flows.  Since Taylor's pioneering work, \cite{Taylor_74}, which invoked the conservation of magnetic helicity in the relaxation of toroidal discharges, an extensive literature has arisen on this topic, much of it reviewed by \cite{Taylor_86}. 

Taylor's original theory, \cite{Taylor_74}, applied globally throughout the plasma with very few adjustable parameters, and was remarkable for its success in modelling toroidal field reversal and helical bifurcation in the highly turbulent reversed-field pinch, Zeta. However, this simplicity restricts its ability to model better-confined  axisymmetric plasmas, \cite{Bhattacharjee_Dewar_82}, such as tokamaks or more modern reversed-field pinches, whose modelling needs more constraints to increase flexibility in matching observed profiles. It is even less adequate for modelling non-axisymmetric systems, such as tokamaks [see e.g. Figs. 7 and 8 of \cite{Hudson_etal_12b}] or reversed-field pinches, \cite{Dennis_etal_13}, with weakly broken symmetry; or stellarators designed from the outset to be non-axisymmetric. 

As will be explained further in Sec.~\ref{sec:Generalizations} we are led to generalize Taylor's theory by replacing the smooth constraints of \cite{Bhattacharjee_Dewar_82} with \emph{singular constraints}---macroscopic current sheets, $\Gamma_{i,j}$, which partition the plasma into multiple relaxation regions $\Omega_i$. [These current sheets may be thought of as thin, flexible sheets of ideal plasma, within which all the IMHD invariants apply, corresponding to the ``singular Casimir elements'' of \cite{Yoshida_Dewar_12}.] This leads to a further dynamical generalization, \emph{multi-region relaxed MHD} (MRxMHD), which we anticipate will have a number of applications in modelling toroidal confinement devices in which three-dimensional geometry effects are important. Current-sheet and relaxation theory applications are also ubiquitous in astrophysical contexts, \cite{Parker_94}, though one has to deal in these applications with the constraint of line-tying, not covered in this paper. 

A new numerical approach to calculating plasma equilibria, using a static version of MRxMHD, has been implemented in a code, SPEC, \cite{Hudson_etal_12b}, particularly useful in non-axisymmetric toroidal fusion confinement systems when most flux surfaces are destroyed by field-line chaos but also capable of reproducing ideal-MHD calculations in systems with a continuous symmetry by using a large number of nested annular toroidal subregions,  \cite{Dennis_Hudson_Dewar_Hole_13}. This static formulation has also recently been used to explore MHD singularities at resonant magnetic surfaces, \cite{Loizu_etal_15a,Loizu_etal_15b}.  Interestingly, our MRxMHD formulation has some striking similarities to an early computational ``water bag'' approach proposed, but apparently not developed further, by \cite{Potter_76}. 

This paper attempts to construct a general formal framework for MRxMHD from first principles in a pedagogic manner, citing related historical and recent work where possible, indicating the scope of MRxMHD, and setting the stage for further development and application. Taylor relaxation and helicity conservation are reviewed in Sec.~\ref{sec:FIhel}, with generalizations discussed in Sec.~\ref{sec:Generalizations}. 
In Secs.~\ref{sec:Kinematics} and \ref{sec:Variations} we review standard fluid Lagrangian variational results, using a notational framework that is formally precise for use in further work. 
The Lagrangian framework for MRxMHD is developed in  Sec.~\ref{sec:Lagrangian}, and MRxMHD dynamics is derived from Hamilton's Principle (of stationary action) in Secs.~\ref{sec:volvar} and \ref{sec:surfvar}, with constraints corresponding to our chosen subset of IMHD invariants. Possible further developments are suggested in the Conclusion. Appendix \ref{sec:thermo} reviews the simple thermodynamics used in MRxMHD and Appendix \ref{sec:Abdy} reviews the boundary condition for the vector potential.

\section{Plasma relaxation}\label{sec:Relaxation}
\subsection{Helicity conservation and Taylor relaxation}\label{sec:FIhel}

The conservation of magnetic fluxes threading cuts through $\Omega$ (or $\Omega_i$ in MRxMHD) that leave it topologically connected follows simply from $\divv\vrm{B} = 0$ and the boundary condition \eqref{eq:tangential}
\emph{everywhere} on $\partial\Omega$ (because of the no-gaps assumption mentioned in the Introduction). Thus, while these fluxes are IMHD invariants, they must also be invariants in any physical model --- flux conservation is said to be completely \emph{robust}.

As it is conserved  even under reconnection [in the limit as resistivity approaches zero, see e.g. eq.~$\!(20)$ ff. of \cite{Jensen_Chu_84}] and has a robust topological interpretation [see e.g. \cite{Berger_99} for a heuristic review or \cite{Arnold_Khesin_98} for a more mathematical treatment] the most robust of the remaining IMHD invariants is widely accepted to be the \emph{magnetic helicity} $2\muSI K_{\Omega}$, where, \cite{Bhattacharjee_Dewar_82}, we define the invariant $K_{\Omega}$ as
\begin{equation}\label{eq:Helicity}
	K_{\Omega} \equiv \int_{\Omega} \frac{\vrm{A}\dotv\vrm{B}}{2\muSI} \, \d V \;,
\end{equation}
with $\vrm{A}$ a vector potential giving $\vrm{B} = \curl\vrm{A}$ and $\d V = \d^3x$ the volume element. Because $K$ has one less gradient of $\vrm{A}$ than the magnetic energy,
\begin{equation}\label{eq:WBdef}
	W^B_{\Omega} \equiv \int_{\Omega} \frac{B^2}{2\muSI} \d V \;,
\end{equation}
it can also be argued, see e.g. Sec.~I.C of \cite{Taylor_86}, that, in a weakly resistive plasma with small-scale turbulent fluctuations, $K$ decays slower with time than $W^B$. 

Other general ideal invariants can be related, by Noether's theorems, to some symmetries in appropriate parameterizations of field variables in the IMHD action (for example, the cross helicity $\int_\Omega\!\vrm{v}\dotv\vrm{B}\d V$ pertains to a relabelling symmetry in the Lagrangian representation of the fields), \cite{Salmon_88,Padhye_Morrison_96a,Padhye_Morrison_96b,Webb_Zank_07,Webb_etal_2014_I,Webb_etal_2014_II,Araki_15}. However, in the spirit of Taylor relaxation we choose the minimal set required to obtain a non-trivial solution and thus keep only $K_{\Omega}$ as the only non-holonomic constraint involving $\vrm{v}$ or $\vrm{B}$.

It is readily shown, using the tangential-$\vrm{B}$  condition \eqref{eq:tangential},
that $K_{\Omega}$ is invariant under gauge transformations $\vrm{A} \mapsto \vrm{A} + \grad\chi$ as long as $\chi$ is single-valued (implying conservation of line integrals $\oint_{\partial\Omega}\!\!\vrm{A}\dotv \d\vrm{l}$ around loops on the boundary, which, by Stokes' theorem, is equivalent to the above-mentioned conservation of magnetic fluxes). As tangential $\vrm{B}$ is to be a holonomic constraint rather than a natural boundary condition, we do not treat $\vrm{A}$ on the boundary as freely variable and can constrain $\chi$. Thus we do not need to use either the Bevir--Gray [subtraction of products of toroidal and poloidal loop integrals, \cite{Bevir_Gray_82}] or relative helicity (subtraction of vacuum-field helicity, \cite{Jensen_Chu_84}) modifications of the helicity, the latter fact also implying there is no physical necessity to decompose the magnetic field into a vacuum (harmonic) and a plasma-current-generated component [though it may still be useful conceptually and mathematically, \cite{Yoshida_Giga_90,Yoshida_Dewar_12}].

The Woltjer--Taylor variational principle [originally proposed, though with less physical motivation, by \cite{Woltjer_58a}] is that the final relaxed state is that which minimizes the magnetic energy, \eqref{eq:WBdef} [the negative of which occurs in \eqref{eq:L}], under the magnetic helicity constraint, implemented by minimizing $W^B_{\Omega} - \mu K_{\Omega}$ under variations of the magnetic vector potential $\vrm{A}$, $\mu$ being a Lagrange multiplier%
\footnote{Such Beltrami constants $\mu$ have dimensions of inverse length and are not to be confused with the vacuum permeability constant $\muSI$ used in SI units.}. 
The resulting Euler--Lagrange equation is the \emph{Beltrami equation},
\begin{equation}\label{eq:Beltrami}
	\curl\vrm{B} = \mu \vrm{B} \;.
\end{equation}
This describes a force-free field, i.e.  one with the current $\vrm{j} = \curl\vrm{B}/\muSI$ parallel to $\vrm{B}$, implying $\grad p = 0$ globally in an equilibrium plasma. Thus this single-region relaxation principle describes only plasmas with no thermal confinement. 

\subsection{Generalization of Taylor Relaxation}
\label{sec:Generalizations}

As the Woltjer--Taylor variational principle is not explicitly based on knowledge of the detailed sub-macroscopic physics leading to relaxation and self-organization, its applicability to modelling a given system can only be justified empirically.%
\footnote{This indeed is also true of ideal MHD, which is typically applied in fusion physics well beyond the validity of the approximations required for using it to describe fusion plasmas [see e.g. Sec.~II.H  of \cite{Freidberg_82}]. In particular, particle mean free paths parallel to magnetic field lines are \emph{not} short in high-temperature plasmas, so modifications of IMHD that distinguish parallel and perpendicular physics have long been sought in order to extend its applicabilty [e.g.. the collisionless MHD of \cite{Freidberg_87}].}
We have already remarked in the Introduction that in fact Taylor relaxation theory in its original form is too simple to apply to modern fusion devices, but that it can be extended naturally by supplementing magnetic helicity with further global ideal invariants, thus preserving much of its simplicity but increasing its flexibility in applications to modelling fusion plasmas. 
 
The fundamental basis of our generalized MHD relaxation principles is the requirement that the states they describe be a subset of the states allowed within ideal MHD. This is ensured, \cite{Bhattacharjee_Dewar_82,Dewar_etal_08}, by using only constraints from a subset of those implied by IMHD. We take this formal criterion as the paramount principle for constructing consistent modifications of IMHD, regarding considerations of possible sub-scale physics that might lead to breaking of some ideal invariants and not others only as an heuristic guide in choosing an appropriate subset of IMHD constraints. For instance, in choosing relaxation subregions in which to apply the Woltjer--Taylor variational principle we do \emph{not} necessarily assume the magnetic field is wholly or partially chaotic, though Beltrami solutions can accommodate such cases, \cite{Dombre_etal_1986}. Justification for the choice of constraints must ultimately be empirical, by comparison either with experiment or \emph{ab initio} simulations.

A generalization of the Taylor relaxation idea by increasing the number of constraints was proposed by \cite{Bhattacharjee_Dewar_82}, but the smooth IMHD invariants chosen then are not well-defined in a nonintegrable magnetic field with islands and chaotic regions. More recently,  \cite{Hudson_Hole_Dewar_07,Dewar_etal_08}, generalizations of Taylor relaxation theory were proposed, based on the assumption that Taylor-relaxed plasma can coexist with \emph{current sheets} that act as transport barriers partitioning $\Omega$ into multiple regions $\Omega_i$, invariant under field-line flow, \cite{Hudson_etal_12b}. 
To describe this approach we have introduced the terminologies RxMHD when $\Omega$ is not partitioned, and MRxMHD when it is, the D (for ``dynamics'') being justified below. The $\delta$-function currents in MRxMHD are compatible with IMHD so they may be regarded as singular alternatives to the smooth IMHD constraints of \cite{Bhattacharjee_Dewar_82}. As in \cite{Bhattacharjee_Dewar_82} we also introduce entropy constraints to allow a nontrivial pressure profile and retain the non-singular magnetic helicity invariant(s), \eqref{eq:Helicity}, but separately conserved in each MRxMHD subregion.

Our development of MRxMHD is implicitly based on the multiple-timescale scenario sketched below 
(a conceptual framework motivating the formal development---whether there are situations where it approximates physical reality remains to be explored in further work):

\begin{enumerate}[1.\,]
\item A fast \emph{relaxation} timescale 
during which all but a finite number of IMHD constraints are broken through thermal diffusion and micro-reconnection events (associated with unspecified mechanisms like micro-tearing turbulence, high-order resonant structures%
\footnote{We use the terminology ``resonant structure'' to denote a family of closed field lines, whose \emph{order} is the number of toroidal rotations they make before they close. A resonant toroidal flux surface is a special case of such a structure, but transient resonant structures may form due to Sweet--Parker reconnection, \cite{Parker_94}, initiated at initially isolated hyperbolic closed field lines such as the ``X points'' of magnetic islands. 
While numerical evidence, \cite{Longcope_Strauss_93,Cordoba_Marliani_00}, that \emph{strictly} $\delta$-function current sheets can form in finite time is not conclusive, it is a reasonable postulate in our coarse-grained, long-reconnection-timescale MRxMHD model.} 
and field-line chaos). On this timescale the system self-organizes into multiple Taylor states in disjoint subregions $\Omega_i$ with non-disjoint boundaries $\partial\Omega_i$ (geometrically fixed on this timescale due to plasma inertia) supporting current sheets on their common interfaces $\Gamma_{i,j}$ [recent simulations by \cite{Smiet_etal_15} give some support for this scenario]. The tangential-$\vrm{B}$  boundary condition, \eqref{eq:tangential}
is satisfied on both sides of these interfaces, but in general $\vrm{B}$ suffers a tangential discontinuity across them. 
[To represent discontinuities across an interface $\Gamma_{i,j}$, we distinguish its inward and outward faces $\Gamma^{\mp}_{i,j}$ by the superscripts $-$ and $+$, respectively, following \cite{McGann_Hudson_Dewar_vonNessi_10}.]\\

\item
An intermediate \emph{dynamical} timescale (the timescale treated in this paper) during which the plasma, including a 
number of embedded current sheets separating sub-regions within which magnetic helicity, magnetic fluxes, mass, and entropy are conserved, evolves adiabatically with respect to the relaxation timescale as the geometric shapes of the boundaries $\partial\Omega_i$ evolve dynamically from their initial conditions, and possibly in response to external forcing from the ``switching on'' of boundary ripple, \cite{Hahm_Kulsrud_85,Dewar_Bhattacharjee_Kulsrud_Wright_13,Comisso_Grasso_Waelbroeck_15a,Comisso_Grasso_Waelbroeck_15b}).
Low-order resonant structures within the plasma that are excited, \cite{Boozer_Pomphrey_10,White_13},  by geometric change resist the formation of magnetic islands by developing \emph{shielding} current sheets.\\

\item A long \emph{reconnection} timescale on which plasma and magnetic flux leaks and mixes between sub-regions through weak spots in the current sheets $\Gamma_{i,j}$, violating the mass and flux isolation of the sub-regions assumed in MRxMHD and also violating entropy conservation. New subregions may form, changing the topological structure of the system. Phenomena on this timescale are not treated in this paper.\\

\end{enumerate}

\begin{figure}
\begin{center}
\begin{tabular}{cc}
	\includegraphics[width=0.2\textwidth]{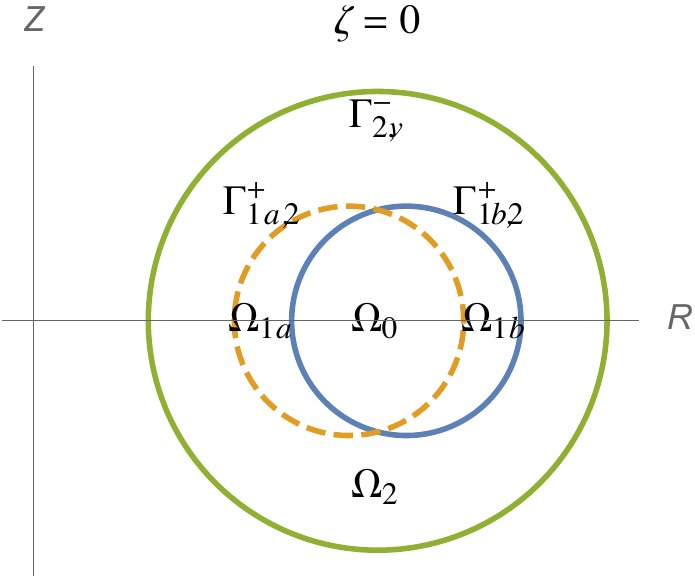}
      &
	\includegraphics[width=0.2\textwidth]{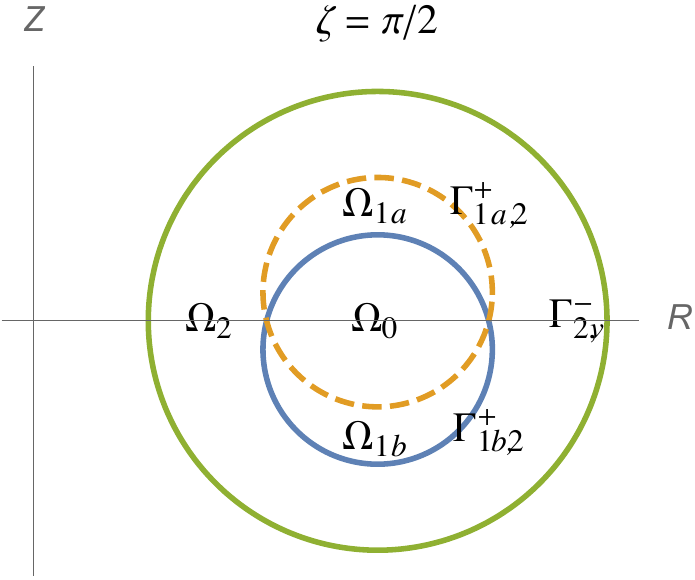}\\
	\includegraphics[width=0.2\textwidth]{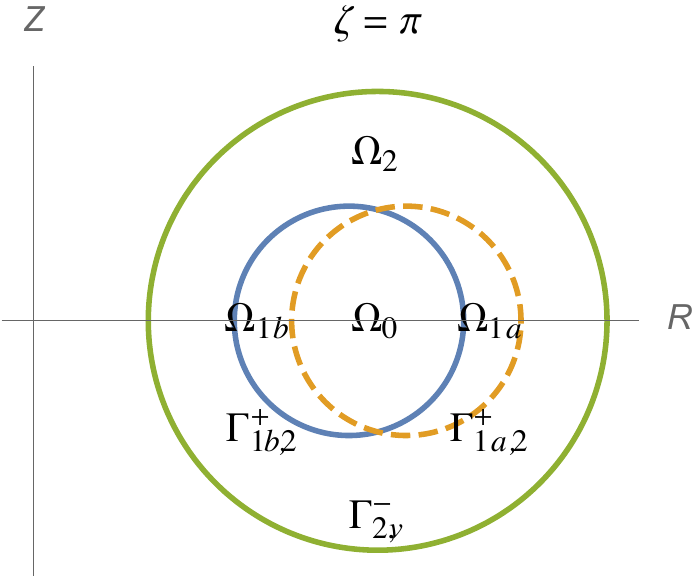}
      &
	\includegraphics[width=0.2\textwidth]{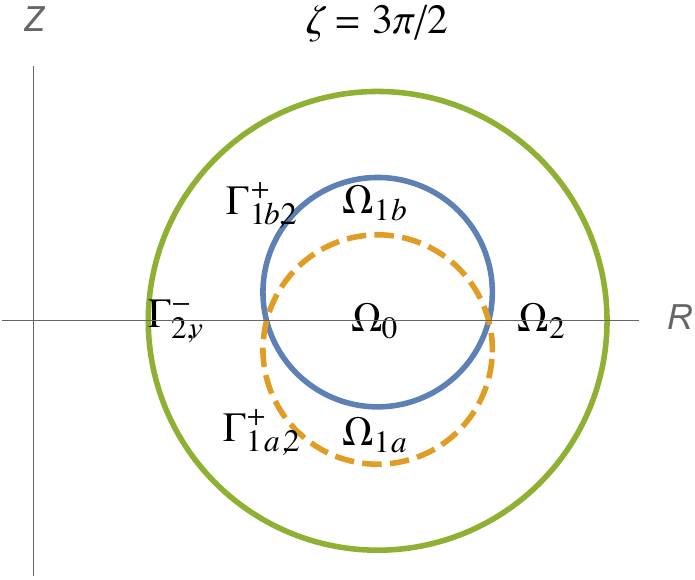}
\end{tabular}
\caption{Four sections of the regions shown in Fig.~\ref{fig:Topologies} at toroidal angles $\zeta = 0$, $\pi/2$, $\pi$ and $3\pi/2$, showing helical rotation of the islands $\Omega_{1a}$ and $\Omega_{1b}$ about the central region $\Omega_0$. The outer region, $\Omega_2$, is an annular genus-2 toroid bounded by $\Gamma_{1a,2}^+$, $\Gamma_{1b,2}^+$ and $\Gamma_{2,\rm v}^-$, the plasma-vacuum interface (green torus).
\label{fig:topolSecs}}
\end{center}
\end{figure}

To illustrate these concepts, a multi-region case of interest is shown schematically in Figs.~\ref{fig:Topologies} and \ref{fig:topolSecs}: an island chain $\{\Omega_{1a},\Omega_{1b}\}$ twisting, with linking number 2, \cite{Berger_99}, around a simple toroid $\Omega_0$, both contained within an annular toroid $\Omega_2$. This illustrates the flexibility of the MRxMHD partition---it is  not limited to simply nested tori as it can include island or plasmoid chains. (This requires a more complicated labelling system for the relaxation regions and current sheets, but if we limit ourselves to primary island chains we can still use a sequential numbering system and indicate the component islands and separatrices of the chain using the lettering scheme illustrated here.)

In single-region RxMHD the topology of $\Omega$ determines the number, $\nu$%
\footnote{The topologically \emph{genus} (or first Betti number) $\nu$ is the number of cuts required to make $\Omega$ simply connected. E.g. in a \emph{simple toroid} (cf. $\Omega_0$ in Figs.~\ref{fig:Topologies} and \ref{fig:topolSecs}), $\nu = 1$, as a toroidal cut leaves it simply connected.
},
of independent fluxes $\Phi^l$, $1 \leq l \leq \nu$, that must be specified for uniqueness of a harmonic (vacuum) field $\vrm{B}_{\rm H}$ solution such that $\curl\vrm{B}_{\rm H} = 0$ in $\Omega$, $\vrm{n}\dotv\vrm{B}_{\rm H} = 0$ on $\partial\Omega$. For $\mu$ not an eigenvalue%
\footnote{In the present context this means $\mu_i$ \emph{below} the lowest Beltrami eigenvalue in each $\Omega_i$, else the plasma would be unstable to local tearing instability, in which case $\Omega_i$ should be partitioned further to raise the minimum eigenvalue, \cite{Dennis_etal_13}.} 
of the Beltrami equation, \eqref{eq:Beltrami}, with homogenous boundary conditions [see e.g. Sec.~IV of \cite{Taylor_86}], specifying the $\nu$ fluxes $\Phi^l$ also specifies the Beltrami field uniquely, \cite{Yoshida_Giga_90}. Similarly, in MRxMHD we need to determine the genus $\nu_i$ of each relaxation sub-region and specify its fluxes $\Phi_i^l$, which are invariant under relaxation.

In the case of an \emph{annular toroid} $\Omega_i$ (e.g. $\Omega_2$ in Figs.~\ref{fig:Topologies} and \ref{fig:topolSecs}), $\partial\Omega_i$ consists of two disjoint tori (e.g. $\Gamma_{1a}^+\cup\Gamma_{1b}^+$ and $\Gamma_2^-$). A standard single-handled torus (i.e. with one hole) can be covered by a single coordinate chart, typically using a \emph{poloidal angle} $\theta$ and a \emph{toroidal angle} $\zeta$.  Assuming its boundaries to be two such standard tori, the genus of an annular toroid is $\nu = 2$, as both toroidal and poloidal cuts are required to make it simply connected.

Higher-genus cases might arise in toroidal confinement when treating doublet/multi-pinch, \cite{Taylor_86}, or bundle divertor, \cite{Stott_Wilson_Gibson_77}, configurations. In the bundle divertor case, the plasma-vacuum boundary $\partial\Omega$ is a two handled torus, which cannot be described by a single toroidal-poloidal coordinate system but must instead be partitioned into two separate patches (coordinate charts). We shall not consider such exotic cases further in this paper.
 
The conservation of the $\Phi_i^l$ implies boundary constraints on the vector potential $\vrm{A}$: From Stokes' theorem
\begin{equation}\label{eq:LoopFlux}
	\Phi_i^l \equiv \int_{\sigma_i^l}\vrm{B}\dotv\vrm{n}\,\d S = \oint_{\gamma_i^l}\!\!\vrm{A}\dotv \d\vrm{l}
\end{equation}
where $\vrm{n}$ is the unit normal at a point on the $l$th topologically distinct surface of section $\sigma_i^l$ cutting $\Omega_i$ and $\gamma_i^l = \partial\sigma_i^l \in \partial\Omega_i$ is a loop around the boundary of $\sigma_i^l$, with direction with respect to that of $\vrm{n}$ given by the right-hand rule.

As the loops $\gamma_i^l$ lie on the boundary $\partial\Omega_i$, which is composed of the current-sheet interfaces $\Gamma_{i,j}$ where there are $\delta$-function currents causing tangential discontinuities in $\vrm{B}$, one might think the values of the loop integrals would depend on whether the loops traverse the inner or outer faces of the $\Gamma_{i,j}$.  However, it does not matter which faces are used as $\vrm{B}$ remains finite within the current sheet---hence, being of infinitesimal width, a current sheet contains only infinitesimal flux. However, this continuity of its loop integrals does not necessarily mean $\vrm{A}$ itself is continuous across current sheets, as a discontinuous gauge term $\grad\chi$ does not affect the loop integrals (provided $\chi$ is single valued). This freedom allows a coordinate-dependent gauge to be used in each relaxation region, as in the SPEC code, \cite{Hudson_etal_12b}.

It is important to recognize that the fluxes $\Phi_i^l$ depend only on magnetic fields within $\Omega_i$, so their individual conservation constrains line integrals of $\vrm{A}$ only around loops that enclose the plasma within $\Omega_i$. However, taking into account the conservation of \emph{all} the $\Phi_i^l$ leaves only the toroidal line integral $\oint_{\partial\Omega}^{\rm tor}\!\!\vrm{A}\dotv \d\vrm{l}$ on the plasma-vacuum interface unconstrained. This represents the external poloidal flux threading the hole in the torus $\partial\Omega$. While this flux is arbitrary as far as the physics of the plasma within $\Omega$ is concerned, it is still \emph{conserved} because we are assuming the wall acts as a superconducting shell which traps the external poloidal flux threading it.

\section{Lagrangian formulation via Hamilton's Principle}\label{sec:LagHam}
\subsection{Lagrangian and Eulerian fluid kinematics}\label{sec:Kinematics}

Central to the Lagrangian approach to fluid mechanics is the concept of \emph{fluid elements}, whose motions with respect to time $t$ through a 3-dimensional Cartesian frame (points in which we designate by the vector $\vrm{x} \equiv x\vrm{e}_x + y\vrm{e}_y + z\vrm{e}_z$), are described by a family of feasible trajectories (pathlines respecting the constraints) $\vrm{x}^t = \vrm{r}^t(\vrm{x}_0)$, labelled by $\vrm{x}_0$, the initial positions of fluid elements at an arbitrary time $t=t_0$. The fluid elements are advected by the Eulerian velocity vector field $\vrm{v}(\vrm{x},t)$ to their positions at arbitrary time $t$ through the \emph{time evolution function} $\vrm{r}^t(\vrm{x})$ defined as the integral of the following equation and initial condition
\begin{equation}\label{eq:tAdvPt}
	\frac{\d\vrm{r}^t(\vrm{x})}{\d t} \equiv \vrm{v}\!\left(\vrm{r}^t(\vrm{x}),t\right)\;,
	\quad \vrm{r}^{t_0}(\vrm{x}) \equiv \vrm{x}
\end{equation}
for all $\vrm{x}$ in the domain of interest.
(Note that this makes $\vrm{r}^t$ implicitly a function of $t_0$, which we can make explicit when needed using the notation $\vrm{r}^t(\vrm{x}|t_0)$, constant parameters such as $t_0$ being listed after the vertical bar $|$.)

Where an argument (other than $t$) is not specified, we treat $\vrm{r}^t$ as a map $\mathbb{R}^3\to\mathbb{R}^3$ (i.e. a 3-vector function of 3-vectors), but often it is necessary to recognize that it is also a \emph{functional} of $\vrm{v}(\vrm{x},t)$, which will be indicated explicitly when required using the notation $\vrm{r}^t[\vrm{v}](\vrm{x})$. This provides a very flexible notation that can be adapted for generating other maps.

Suppose $\vrm{v}$ is also a function of some time-independent parameter [say $s$, denoted by $\vrm{v}(\vrm{x},t|s)$] then $\vrm{r}^t$ will also be a function of $s$, denoted $\vrm{r}^t(\vrm{x}|s,t_0)$. Suppose further that $\vrm{x}$ lies on an arbitrary curve $\vrm{x} = \vrm{f}(s)$ at time $t_0$ and denote the resulting family of trajectories by $\vrm{R}^t(s) \equiv \vrm{r}^t(\vrm{f}(s)|s,t_0)$. Differentiating both sides of the equation of motion in \eqref{eq:tAdvPt} with respect to $s$, we have
\begin{equation}\label{eq:tAdvParamDeriv}
	\frac{\d}{\d t}\frac{\d\vrm{R}^t}{\d s} = \frac{\d\vrm{R}^t}{\d s}\dotv\grad\vrm{v}(\vrm{R}^t,t) + \partial_s\vrm{v}(\vrm{R}^t,t) \;,
\end{equation}
where $\partial_s$ means the partial derivative with respect to $s$ and $\grad\vrm{v}$ denotes $\grad_{\vrm{x}}\vrm{v}(\vrm{x},t)$.

We now use the special case of the above result where $\vrm{v}$ is independent of $s$ to build up some useful differential-geometric evolution results. First, denoting an infinitesimal line element advected by the fluid by $\d\vrm{l}^t \equiv \d\vrm{R}^t$, \eqref{eq:tAdvParamDeriv} immediately gives
\begin{equation}\label{eq:tAdvLine}
	\frac{\d}{\d t}\d\vrm{l}^t = \d\vrm{l}^t\dotv\grad\vrm{v} \;.
\end{equation}

When acting on all points in a region $\Omega$, the evolution function $\vrm{r}^t$ defines the \emph{Lagrangian map}, mapping an initial  region $\Omega_0$ onto its image $\Omega^t$. From \eqref{eq:tAdvLine} applied to the sides of an infinitesimal rhomboid within $\Omega^t$ the advection equation for infinitesimal volumes $\d V^t$ is found to be
\begin{equation}\label{eq:tAdvVol}
	\frac{\d}{\d t}\d V^t = (\divv\vrm{v})\d V^t  \,,
\end{equation}
which is equivalent to the evolution equation, $(\d/\d t)J(t) = (\divv\vrm{v})J(t)$, for the \emph{Jacobian}, $J(t) = \d V^t/\d V_0 \equiv \partial(x^t,y^t,z^t)/\partial(x_0,y_0,z_0)$, of the transformation from initial coordinates of fluid elements to the corresponding coordinates at time $t$.

Likewise the boundary $\partial\Omega_0$ maps onto $\partial\Omega^t$, within which area elements $\d\vrm{S}^t \equiv \vrm{n}^t dS^t$ advect according to
\begin{equation}\label{eq:tAdvSurf}
	\frac{\d}{\d t}\d\vrm{S}^t = (\divv\vrm{v}) \d\vrm{S}^t - (\grad\vrm{v})\dotv \d\vrm{S}^t
	\;.
\end{equation}

Representing a Lagrangian map as a dynamical \emph{flow}, induced by an Eulerian velocity field, allows connection to be established with the modern Lie algebra approach to fluid dynamics, \cite{Arnold_Khesin_98}. However, although Lie operator methods are useful in Hamiltonian perturbation theory, \cite{Dewar_76b}, the simpler Lagrangian-based variational approach used in this paper avoids the need for most of this abstract machinery. (Likewise for abstract differential geometry.) Nevertheless we shall find the extension of our Eulerian--Lagrangian mapping notation to include flows other than time evolution makes for a compact notation, and, being semi-Eulerian, leads to a more familiar form for the perturbation expansion of the Lagrangian than the strictly Lagrangian approach, \cite{Dewar_70}.

Application of Hamilton's Principle requires us to vary trial fluid element pathlines to find the Euler--Lagrange equations that determine which such pathlines are actually physical. Thus we introduce a new flow that maps the position vectors of fluid elements from their unvaried positions at each time $t$ to their varied positions at the same time by using a flow analogous to the Lagrangian map defined in \eqref{eq:tAdvPt}, but with $t$ replaced by a dimensionless \emph{variation parameter} $\epsilon$ (typically small), and with the \emph{variational map} generated by a \emph{variational velocity} $\bm{\nu}$. 
Thus, in eqs.~$\!$(\ref{eq:tAdvPt}--\ref{eq:tAdvSurf}) replace $t$ with $\epsilon$, the initial time $t_0$ with 0, the time evolution flow $\vrm{r}^t[\vrm{v}]$ with the $\epsilon$-\emph{flow} $\vrm{r}^\epsilon[\bm{\nu}]$,  and the velocity $\vrm{v}(\vrm{x},t)$ with $\bm{\nu}(\vrm{x},\epsilon|t)$, to give
\begin{equation}\label{eq:epsAdvPt}
	\frac{\d\vrm{r}^\epsilon(\vrm{x})}{\d \epsilon} \equiv \bm{\nu}\left(\vrm{r}^\epsilon(\vrm{x}),\epsilon\right)\;,
	\quad \vrm{r}^{\epsilon=0}(\vrm{x}) \equiv \vrm{x} \;,
\end{equation}
with the parametric time-dependence $(\ldots|t)$ now left implicit.

The varied Lagrangian map $\vrm{r}^t[\vrm{v}_\sim]$ is now found by composing the unvaried Lagrangian map with the variational map,
\begin{equation}\label{eq:CompositeEvol}
	\vrm{r}^t[\vrm{v}_\sim]
	\equiv \vrm{r}^{\epsilon}[\bm{\nu}]\comp\vrm{r}^t[\vrm{v}] \;,
\end{equation}
where $\comp$ denotes composition of functions: $f\comp g(x) \equiv f(g(x))$ and subscript $\sim$ denotes a varied quantity. This implicitly defines the varied velocity field $\vrm{v}_\sim$, which is to be found in terms of the unvaried position of a representative fluid element $\vrm{x}^t \equiv \vrm{r}^t(\vrm{x}_0)$ as the total time derivative of the varied position, $\vrm{x}_\sim^t \equiv \vrm{r}^t[\vrm{v}_\sim](\vrm{x}_0) \equiv \vrm{r}^{\epsilon}(\vrm{x}^t|t)$, as
\begin{equation}\label{eq:vtildeGen} 
	\vrm{v}_\sim (\vrm{x}_\sim,t)
	 =  
	  \mathrm{D}_t \, \vrm{x}_\sim
		\;,
\end{equation}
where $\vrm{x}_\sim$ here denotes the Eulerian representation of the varied position, $\vrm{x}_\sim(\vrm{x},t) \equiv \vrm{r}^{\epsilon}(\vrm{x}|t)$, and $\mathrm{D}_t \equiv \partial_t + \vrm{v}(\vrm{x},t)\dotv\grad$ is the advective derivative.

In MRxMHD we also need to consider the case of fluid elements on the common interfaces (current sheets) $\Gamma_{i,j} = \partial\Omega_i\cap\partial\Omega_j$ separating subregions $\Omega_i$ and $\Omega_j$. The shape of the interface, which we represent as the level surface $f_{(i,j)} = 0$ of an appropriate smooth function $f(\vrm{x})$ changing monotonically across the surface, is not known \emph{a priori} so must be subject to variation in applying Hamilton's Principle. Thus we must introduce the variation parameter $\epsilon$ in representing $\Gamma_{i,j}$ geometrically: $f_{(i,j)}(\vrm{x}|t,\epsilon) = \pm 0$. We use the notation $\vrm{x}_{\sim\pm}^t$ to distinguish which side of the interface a varied fluid element is on: $f_{(i,j)}(\vrm{x}_{\sim\pm}^t|t,\epsilon) = \pm 0$. Taking the total derivative of left and right sides of this expression with respect to $t$, on both sides of the interface, we find $\vrm{n}\dotv\jump{\vrm{v}_\sim} = 0$, $\vrm{n} \equiv \grad\! f_{(i,j)}/|\grad\! f_{(i,j)}|$ being the unit normal and $\jump{\cdot}$ denoting the \emph{jump} in a quantity as the evaluation point crosses the interface (so $\jump{\partial_t f_{(i,j)}(\vrm{x}|t,\epsilon)} = 0$, as $f_{(i,j)}$ is assumed smooth). Similarly, total differentiation with respect to $\epsilon$ gives
\begin{equation}\label{eq:innerrconstraint}
	\vrm{n}\dotv\jump{\bm{\nu}} = 0,\quad \vrm{x}\in\Gamma^{\pm}_{i,j} \;.
\end{equation}
This states that the normal component of $\bm{\nu}$ is constrained to be continuous across the interface, but otherwise it is unconstrained. Rather, $\Gamma_{i,j}$ is advected with $\bm{\nu}$ during variations at constant $t$, just as it is under time evolution.

An exception is the case where the plasma is confined by a prescribed, though possibly time-dependent, boundary ``wall'' (w). This is the special case that $f_{(i,j=\rm w)}$ is not a function of $\epsilon$, so one obtains the constraint $\vrm{n}\dotv\bm{\nu} = 0$ at constant $t$ on $\Gamma^-_{(i,\rm w)}$, with the tangential components unconstrained.

So far we have treated $\epsilon$ as a finite parameter, on a par with $t$. However, in this paper we use the variational transformation only for calculating the first variation of the action in Hamilton's Principle, so we need the variational map only to linear order: $\vrm{r}^\epsilon[\bm{\nu}](\vrm{x}) = \vrm{x} + \epsilon\bm{\nu}(\vrm{x},0|t) + O(\epsilon^2)$. 
Thus, defining the Lagrangian variation in position $\Delta\vrm{x}$ through $\vrm{x}_\sim = \vrm{x}  +\epsilon\Delta\vrm{x}(\vrm{x},t) + O(\epsilon^2)$, we have $\Delta\vrm{x}(\vrm{x},t) = \bm{\nu}(\vrm{x},0|t)$. Similarly, defining the Lagrangian variation in velocity, $\Delta\vrm{v}$, to be such that $\vrm{v}_\sim (\vrm{x}_\sim,t) = \vrm{v}(\vrm{x},t) + \epsilon\Delta\vrm{v} (\vrm{x},t)  + O(\epsilon^2)$, we have, from \eqref{eq:vtildeGen},
\begin{equation}\label{eq:vtilde} 
	\Delta\vrm{v} (\vrm{x},t) =  \mathrm{D}_t\,\Delta\vrm{x}(\vrm{x},t)
		\;.
\end{equation}

In the above we have followed \cite{Newcomb_62} in using $\Delta f \equiv \lim_{\epsilon \to 0}[f_\sim(\vrm{x}_\sim,\epsilon|t) - f(\vrm{x},\epsilon|t)]/\epsilon$ to denote the \emph{Lagrangian variation} in an arbitrary field $f$, while the corresponding \emph{Eulerian variation} $\delta\! f$ is defined by $\delta\! f \equiv \lim_{\epsilon \to 0}[f_\sim(\vrm{x},\epsilon|t) - f(\vrm{x},\epsilon|t)]/\epsilon$, i.e. with both the varied and unvaried field evaluated at the same, unvaried, position $\vrm{x}$. Thus $\Delta$ may be regarded as the operator $\lim_{\epsilon \to 0}\d/\d\epsilon$ while $\delta$ is the operator $\lim_{\epsilon \to 0}\partial/\partial\epsilon$.

By definition the two operators are related by $\Delta = \delta + \Delta\vrm{x}\dotv\grad$. Applying both sides to $\vrm{x}$ it is easily verified, as a consistency check, that $\delta\vrm{x} = 0$. Also, both $\delta$ and $\Delta$ being differential operators, the product rule, e.g. $\Delta (fg) = (\Delta f)g + g\Delta f$, and commutation relation
\begin{equation}\label{eq:Commutator}
	\Delta\grad\! f = \grad\Delta f - (\grad\Delta\vrm{x})\dotv\grad\! f
\end{equation}
apply, and correspondingly for $\delta$, where $f(\vrm{x},\epsilon)$ and $g(\vrm{x},\epsilon)$
are arbitrary.


\subsection{Holonomically constrained and free variations}\label{sec:Variations}

In this subsection we extend the general formalism developed above for fluid kinematics to treat variation and perturbation of \emph{fields}, specifically mass density $\rho(\vrm{x},t)$, pressure $p(\vrm{x},t)$, and magnetic vector potential $\vrm{A}(\vrm{x},t)$ [and hence magnetic field $\vrm{B} \equiv \curl\vrm{A}$].

The density (mass conservation) equation is the lowest and most robust one in the hierarchy of moment equations used in deriving fluid models from kinetic theory. In fact freezing mass into fluid elements seems fundamental to any fluid theory, so we build the holonomic mass conservation constraints $\rho_\sim(\vrm{x}_\sim^t,t|\epsilon)\d V_\sim^t  = \rho(\vrm{x}^t,t)\d V^t = \rho(\vrm{x}_0,t_0)\d V_0$ 
into both IMHD [as in \cite{Newcomb_62,Dewar_70}] and into our new MRxMHD formulation. From \eqref{eq:tAdvVol} and its $\epsilon$-flow analogue, these imply
\begin{equation}\label{eq:rhoeqsL}
	\frac{\d\rho}{\d t} = -\rho\divv\vrm{v}, \: \frac{\d\rho_\sim}{\d\epsilon} = -\rho_\sim\divv\bm{\nu} \;,
\end{equation}
or, decomposing $\d/\d t$ as $\partial_t  + \vrm{v}\dotv\grad$, $\d/\d\epsilon$ as $\partial_\epsilon  + \bm{\nu}\dotv\grad$,
\begin{equation}\label{eq:rhoeqsE}
	\frac{\partial\rho}{\partial t} = -\divv(\rho\vrm{v}),  \: \frac{\partial\rho_\sim}{\partial\epsilon} = -\divv(\rho_\sim\bm{\nu}) \;.
\end{equation}
As explained above, the Lagrangian variation in density is $\Delta\rho = \d\rho_\sim/\d\epsilon|_{\epsilon=0}$, and the Eulerian variation is $\delta\rho = \partial\rho_\sim/\partial\epsilon|_{\epsilon=0}$. That is, from eqs.~(\ref{eq:rhoeqsL}) and (\ref{eq:rhoeqsE}),
\begin{equation}\label{eq:mconstraint}
	\Delta\rho = -\rho\divv\Delta\vrm{x} \:\Leftrightarrow\: \delta\rho = -\divv(\rho\Delta\vrm{x}) \;.
\end{equation}

In MRxMHD [unlike IMHD, \cite{Newcomb_62,Dewar_70}] pressure and magnetic field $\curl\vrm{A}$ are not holonomically constrained by the freezing of entropy and flux microscopically into each fluid element but are freely variable, expressed by writing pressure and vector potential as $p(\vrm{x},t|\epsilon)$ and $\vrm{A}(\vrm{x},t|\epsilon)$, with their $\epsilon$-derivatives $\delta p$ and $\delta\vrm{A}$ being arbitrary variations in Hamilton's Principle (as is $\Delta\vrm{x}$).
(This is not of course to say Lagrangian variations of free, i.e. not holonomically constrained, fields do not exist, but rather that their Eulerian variations $\delta$ are \emph{primary}, with their Lagrangian variations being \emph{defined} by the rule $\Delta = \delta + \Delta\vrm{x}\dotv\grad$ above.)
The relation of interface kinematics to the holonomic tangential-$\vrm{B}$ constraint, \eqref{eq:tangential}, 
is developed in Appendix~\ref{sec:Abdy}.

Consider an integral of the form $L = \int_{\Omega}\mathcal{L} \d V$, where $\mathcal{L}$ is a function of various fields to be varied in Hamilton's Principle (in our case $\rho$, $\vrm{v}$, $p$ and $\vrm{A}$), which are thus functions of $\epsilon$.  By differentiating $L$ with respect to $\epsilon$, using the $\epsilon$-flow analogue of \eqref{eq:tAdvVol} (after changing variables to $x^0,y^0,z^0$, where, as in \cite{Frieman_Rotenberg_60}, superscript $0$ refers to $\epsilon = 0$) and integration by parts (Gauss' theorem) we find the convenient identity
\begin{equation}\label{eq:ID1}
	\delta L = \int_{\Omega} \delta \mathcal{L} \, \d V
			+ \int_{\partial\Omega} \mathcal{L}\, \Delta\vrm{x}\dotv\d\vec{S} \;,
\end{equation}
which allows us to use the Eulerian variations $\delta = \partial_\epsilon$ defined above to restrict the occurrence of $\Delta\vrm{x}$ to those fields ($\rho$ and $\vrm{v}$) that are constrained to vary with fluid motions. [An alternative, but equivalent, approach is to transform to $x_0,y_0,z_0$ at the outset, using Lagrangian variations and the $\epsilon$-flow analogue of \eqref{eq:tAdvSurf} to get surface terms.]

In applying \eqref{eq:ID1} to compute the action variation $\delta\mathcal{S} = \int\d t\int_{\Omega^t}\d V\mathcal{L}$, further surface terms will arise after straightforward integrations by parts to remove spatial derivatives of $\Delta\vrm{x}$. However, the integration by parts arising from terms containing time derivatives is more subtle because it involves a change of variables from $x,y,z$ to $x_0,y_0,z_0$ to enable integration by parts, and then a change back to $x,y,z$. In our case the only such term arises from the kinetic energy term, a function of $\vrm{v}$, the time derivative arising from the Eulerian variation $\delta\vrm{v} = \Delta\vrm{v} - \Delta\vrm{x}\dotv\grad\vrm{v} = \mathrm{D}_t\Delta\vrm{x} - \Delta\vrm{x}\dotv\grad\vrm{v}$ by \eqref{eq:vtilde}.

Consider a Lagrangian density of the form $\mathcal{L}(\vrm{v})$. Then, noting that $\Delta\vrm{x}$ is always taken to vanish at the endpoints of the time integration in Hamilton's Principle, and using \eqref{eq:tAdvVol} we find the contribution to the action from $\delta\vrm{v}$,
\begin{equation}\label{eq:ID2}
	\int\d t\int_\Omega\d V\delta\vrm{v}\dotv\frac{\partial\mathcal{L}}{\partial\vrm{v}} \nonumber \\ 
	= -\int\d t\int_{\Omega}\d V\Delta\vrm{x}\dotv\left(\vsf{I}\divv\vrm{v} + \grad\vrm{v}Ê+ \vsf{I}\mathrm{D}_t\right)\dotv\frac{\partial\mathcal{L}}{\partial\vrm{v}} \;,
\end{equation}
which has \emph{no} surface term. 

\subsection{Lagrangian formulation of MRxMHD}\label{sec:Lagrangian}

To derive the dynamics of multi-region relaxed plasmas we use the same Lagrangian density $\mathcal{L}^{\rm MHD}$ as in \eqref{eq:L}, integrating over each subvolume $\Omega _i$ and augmenting with appropriate Lagrange multiplier terms to form the Lagrangian in $\Omega_i$, $L_i$.  We then sum to form the total effective Lagrangian
\begin{equation}\label{eq:LRx}
	L = L _{\rm v} + \!\sum_{i\in\mathcal{R}} L_i 
\;,
\end{equation}
where $\mathcal{R}$ denotes the set of plasma relaxation regions, with their Lagrangians $L_i$ being given by
\begin{eqnarray}\label{eq:Li}
	L_i &= & L^{\rm MHD}_i + \tau_i(S_i - S_{i0}) +\mu_i\left(K_i - K_{i0}\right) \;,
\end{eqnarray}
where $L^{\rm MHD}_i$ is given by the integral of $\mathcal{L}^{\rm MHD}$, defined in \eqref{eq:L}, over $\Omega_i$, the $S_i$ are the entropy invariants ($mS_{\rm en}$ in the notation of Appendix~\ref{sec:thermo}) given by integrating the RHS of \eqref{eq:sm} over $\Omega_i$,
\begin{equation}\label{Sdef}
	S_i \equiv \int_{\Omega_i}\frac{\rho}{\gamma - 1} \ln\left(\kappa\frac{p}{\rho^\gamma}\right) \d V \;,
\end{equation}
and the magnetic helicity invariants $K_i$ are as in \eqref{eq:Helicity}, evaluated over $\Omega_i$. (The constant $\kappa$, required to make the argument of $\ln$ dimensionless, is arbitrary for our purposes but is identified physically in Appendix~\ref{sec:thermo}.)

The Lagrange multipliers $\tau_i$ and $\mu_i$ are constant during variation in Hamilton's Principle (i.e are independent of $\epsilon$) but may change with $t$, as they must be chosen to make $S_i = S_{i0}$ and $K_i = K_{i0}$ during evolution under the Euler--Lagrange equations. The constant reference values $S_{i0}$ and $K_{i0}$ are the respective initial values at $t = t_0$ evaluated over $\Omega_{i0}$, making $L_i = L^{\rm MHD}_i$ when the $\tau_i$ and $\mu_i$ are adjusted to satisfy the conservation conditions above. Subtracting off the constant $K_{i0}$ also makes $L_i$ independent of the constant $\kappa$ (because of the holonomic conservation of $\int\!\rho\, \d V$), so the curious fact that the physical value of $\kappa$, \eqref{eq:kappadef}, involves Planck's constant is irrelevant.

If there is a vacuum region, $\Omega_{\rm v}$, between a plasma-vacuum interface and the wall, then this may be treated similarly, but with $\rho$ set to zero and with the entropy and helicity constraints deleted, leaving the Lagrangian density
\begin{equation}\label{eq:Lv}
	\mathcal{L}_{\rm v} =  - \frac{\vrm{B}\dotv\vrm{B}}{2\muSI} \;.
\end{equation}

After applying eqs.~(\ref{eq:ID1}--\ref{eq:ID2}) and appropriate integrations by parts the variation of the action must be of the general form
\begin{eqnarray}\label{eq:deltaSRx}
	\delta \mathcal{S}&
	= &\!\!\int\!\! \d t\!\!\sum_{i\in\mathcal{R}^+}\!\!\int_{\Omega_i}\!\!\!\!\!\d V\! \left( \delta\vrm{A}\dotv\frac{\delta\mathcal{S}_i}{\delta\vrm{A}}
	+ \delta p\frac{\delta\mathcal{S}_i}{\delta p} + \Delta\vrm{x}\dotv\frac{\delta\mathcal{S}_i}{\delta\vrm{x}}\right)\nonumber\\ 
	&&\mbox{}\!\! + \!\!\int\!\! \d t\!\!\sum_{i\in\mathcal{R}^+}\!\!\int_{\partial\Omega_i}\!\!\!\!\!\!\! \d S \left.\frac{\delta\mathcal{S}_i}{\delta\vrm{x}}\right|_{\partial\Omega_i}\!\!\!\!\!\!\!\!\dotv\,\Delta\vrm{x}  \;,
\end{eqnarray}
where $\mathcal{R}^+$ denotes the set of plasma relaxation regions plus the vacuum region (though note that only the variational derivative $\delta\mathcal{S}_{\rm v}/\delta\vrm{A}$ is non-zero in $\Omega_{\rm v}$). The interval over which the time integral is taken does not need to be specified as variations at the endpoints are taken to vanish in Hamilton's Principle. 

Hamilton's Variational Principle is the statement that the Euler--Lagrange equations following from requiring $\delta \mathcal{S}= 0$ for \emph{all} variations of the independent fields determine these fields physically. In the above variational derivative notation these are the Euler--Lagrange equations $\delta\mathcal{S}_i/\delta\vrm{A} = 0$, $\delta\mathcal{S}_i/\delta p = 0$, $\delta\mathcal{S}_i/\delta\vrm{x} = 0$ within the volumes $\Omega_i$.
To find the surface Euler--Lagrange equations the constraint \eqref{eq:innerrconstraint} needs to be considered, the consequences of which will be discussed in Sec.~\ref{sec:surfvar}.

\subsection{Volume variations}\label{sec:volvar}

Inserting \eqref{eq:Li} in \eqref{eq:LRx}, varying $\mathcal{S}= \int\! L\,\d t$, integrating by parts and comparing with \eqref{eq:deltaSRx}, we identify the variational derivative with respect to the vector potential as
\begin{equation}\label{eq:Avar}
	\frac{\delta\mathcal{S}_i}{\delta\vrm{A}} = -\frac{1}{\muSI}\left(\curl\vrm{B} - \mu_i\vrm{B}\right) \;,
\end{equation}
giving as the corresponding Euler--Lagrange equation, $\delta\mathcal{S}_i/\delta\vrm{A} = 0$, the Beltrami equation \eqref{eq:Beltrami}. In the case of the vacuum Lagrangian $L_{\rm v}$, $\mu_i$ is set to zero, giving the statement that the vacuum field is harmonic, $\curl\vrm{B} = 0$. (While the Euler--Lagrange equation in the vacuum region is the same as the Beltrami equation with $\mu$ set to zero, note that we have actually \emph{deleted} the constraint of constant magnetic helicity in the vacuum region, so the vacuum is not completely equivalent to a currentless plasma.)

The variational derivative with respect to the pressure is
\begin{equation}\label{eq:pvar}
	\frac{\delta\mathcal{S}_i}{\delta p} = \frac{\partial \mathcal{L}_i}{\partial p} = -\frac{1}{\gamma-1}\left(1 - \tau_i\frac{\rho}{p}\right) \;,
\end{equation}
the corresponding Euler--Lagrange equation being
\begin{equation}\label{eq:isoeqstate}
	p = \tau_i \rho \;,
\end{equation}
with $\tau_i$ identified [see \eqref{eq:denspress}] as the \emph{specific temperature} $T_i/m$ in $\Omega_i$, where $m$ is the effective ion mass $m_{\rm i}/Z_{\rm eff}$ ($\tau_i$ is also $C^2_i$, where $C_i$ is the ion sound speed).Ê

Note that, despite using internal energy and entropy densities consistent with a \emph{microscopically} isentropic equation of state, we have only enforced \emph{macrosopic} entropy conservation over a whole subregion, leading to the microscopically \emph{isothermal} equation of state \eqref{eq:isoeqstate} ($\tau_i$ being spatially constant). Rapid equilibration of temperature within $\Omega_i$ is compatible with magnetic surfaces being destroyed within a relaxation region, hence poor local thermal confinement,
but the bounding interfaces are assumed to be thermal transport barriers so the temperature can jump across each interface. 
Thus, while
temperature profiles are restricted to being piece-wise constant they are otherwise arbitrary, enabling the use of MRxMHD to model hot, magnetically confined plasmas [see e.g. \cite{Hudson_etal_12b}].

As already remarked, the Lagrange multipliers $\mu_i$ and $\tau_i$ may change with time to maintain the constancy of their respective constraints. Now that the $\tau_i$ have been identified as temperatures, a simple thought experiment makes it physically clear that this must be so: Suppose the $\rho_i$, and thus the $p_i$, are spatially constant within each $\Omega_i$, then, just as for adiabatically deformed bags of ideal gas,  $\tau_i V_i^{\gamma-1} = \const$---any change in volume $V_i$ leads to a change in temperature $\tau_i$. 

Using \eqref{eq:mconstraint} and \eqref{eq:ID2} we then find the variational derivative with respect to fluid element positions
\begin{eqnarray}\label{eq:rhovar}
	\frac{\delta\mathcal{S}_i}{\delta\vrm{x}} & = & -\partial_t(\rho\vrm{v}) - \divv(\rho\vrm{v}\vrm{v}) - \rho\grad\frac{v^2}{2} + \rho\grad\frac{\partial \mathcal{L}_i}{\partial \rho} \nonumber\\
	 & = &  -\partial_t(\rho\vrm{v}) - \divv(\rho\vrm{v}\vrm{v}) 
	 	 +\frac{\tau_i\rho}{\gamma - 1}\grad\left(\frac{\grad p}{p} -\gamma\frac{\grad \rho}{\rho}\right) \nonumber\\
	 & = & -\partial_t(\rho\vrm{v}) - \divv(\rho\vrm{v}\vrm{v} +  p\, \vsf{I}) \;,
\end{eqnarray}
where the last line follows from the isothermal equation of state \eqref{eq:isoeqstate}. The corresponding Euler--Lagrange equation $\delta\mathcal{S}_i/\delta\vrm{x} = 0$ is the equation of motion for a compressible Euler fluid in momentum conservation form. Note the extraordinary simplicity of this result, with $\vrm{v}$ being decoupled from $\vrm{B}$ as the Beltrami equation \eqref{eq:Beltrami} implies the Lorentz force $\vrm{j}\cross\vrm{B}$ is zero, showing that MRxMHD supports only steady flows and sound waves, with phase velocity $(T/m)^{1/2}$, within the relaxation regions. (However surface waves on the interfaces can involve perturbations of $\vrm{B}$.)

\subsection{Surface variations}\label{sec:surfvar}

It is shown in Appendix~\ref{sec:Abdy} that tangential $\vrm{B}$ at the interfaces $\Gamma_{i,j}$ implies a holonomic constraint on variations of the tangential component, $\vrm{A}_{\rm tgt} \equiv (\vsf{I} - \vrm{n}\vrm{n})\dotv\vrm{A}$, of the vector potential,
\begin{equation}\label{eq:Abc}
	(\delta\!\vrm{A})_{\rm tgt} = (\Delta\vrm{x}\cross\vrm{B} + \grad\delta\chi)_{\rm tgt}  \;,
\end{equation} 
for all $\vrm{x}$ on the interface or boundary, where $\delta\chi$ is an arbitrary single-valued gauge potential. The normal component $\vrm{n}\dotv\delta\!\vrm{A}$ is unconstrained. It is also shown in Appendix~\ref{sec:Abdy} that this constraint implies invariance of line integrals of $\vrm{A}$ around loops on these surfaces, and thus conservation of flux does not need to be imposed as an extra constraint.

Taking into account \eqref{sec:surfvar} the surface variational derivative in \eqref{eq:deltaSRx} is found to be
\begin{equation}\label{eq:surfvar}
	\left.\frac{\delta\mathcal{S}_i}{\delta\vrm{x}}\right|_{\partial\Omega_i} 
	= \left(p + \frac{B^2}{2\mu_0}\right)\vrm{n}_i \;,
\end{equation}
which is the velocity-independent part of the stress tensor, \cite{Dewar_70} dotted with $\vrm{n}_i$.

However, before we can apply this result we need to take into account the fact that a boundary $\partial\Omega_i$ is made up of interfaces between $\Omega_i$ and neighbouring regions $\Omega_j$ (say) across which, by \eqref{eq:innerrconstraint}, the normal components of $\Delta\vrm{x}$ are continuous. Taking into account the outward normals of contiguous regions being oppositely directed, this implies the constraint $\Delta\vrm{x}\dotv\vrm{n}_i = -\Delta\vrm{x}\dotv\vrm{n}_j$ on the common interfaces $\Gamma_{i,j}$. This coupling gives the surface Euler--Lagrange equation, the natural boundary condition between interfaces, as the continuity condition
\begin{equation}\label{eq:normsurfvar}
	\jump{p + \frac{B^2}{2\mu_0}} = 0 \;,
\end{equation}
which is the same as the jump condition for advected discontinuities in ideal MHD [see e.g. Sec.~5.12 of \cite{Hosking_Dewar_15}].
(The tangential components $\Delta\vrm{x}_{\rm tgt}$ are separately variable, but give no natural boundary conditions because they do not appear in $\delta\mathcal{S}$.)

\section{Conclusion}\label{sec:conclude}

We have built a general framework on which to  develop relaxed-MHD dynamics further. Some avenues to be explored are indicated below:
\begin{itemize}
\item
	Stationary states with flow: If one invokes a modified form of the ``imaginary experiment'' of \cite{Kruskal_Kulsrud_58} in which the fictitious friction force acts only on interface movements, and is sufficiently strong so as to allow only movements slow compared with a characteristic sound transit time, then negligible sound wave energy will be excited (see Sec.~\ref{sec:Generalizations}) and the system will relax to a static equilibrium state or one with steady flow. Thus \emph{action} extremization would seem to provide a more physically intuitive framework for variational construction of equilibria with flows than one based on \emph{energy} minimization, which requires an arbitrary angular momentum constraint, \cite{Dennis_Hudson_Dewar_Hole_14a} to keep the kinetic energy from being minimized to zero.
	
It also does not seem necessary to invoke the cross-helicity invariant, \cite{Hameiri_14,Dennis_Hudson_Dewar_Hole_14a}, mentioned in Sec.~\ref{sec:FIhel}.
\item
	Spectral and stability studies with and without flow: By including the kinetic energy in a natural way, our dynamical formulation of MRxMHD provides a physical normalization for the linear growth rates of  instabilities to replace the artificial one derived previously using an energy principle, \cite{Hole_Hudson_Dewar_07,Mills_Hole_Dewar_09}.
	
	The dynamical formulation also suggests performing simulations using the water-bag approach, \cite{Potter_76}, for exploring the nonlinear evolution, and possible saturation, of instabilities. This may, for instance, help resolve the paradox that existence of equilibrium interfaces seems to be contingent on having highly irrational rotational transforms on the domain boundaries $\partial\Omega_i$, \cite{McGann_Hudson_Dewar_vonNessi_10,McGann_13}, but fixing rotational transform is incompatible with the constancy of the helicity invariants in general. Simulation may also be a useful way to explore formation of singularities at resonant surfaces, \cite{Loizu_etal_15a,Loizu_etal_15b}.
\item
	Exploration of the possibility of including reconnection in MRxMHD by allowing a slow leak of flux and plasma through the interfaces.
\end{itemize}

\appendix

\section{Application of ideal-gas thermodynamics to plasmas}
\label{sec:thermo}

The thermodynamics involved in MRxMHD is elementary, being the same as for an ideal gas. However, the expressions used here for the internal energy density $p/(\gamma-1)$ and entropy constraint density $\rho\ln(\kappa p/\rho^{\gamma})/(\gamma-1)$ introduced in \cite{Bhattacharjee_Dewar_82} (an essentially arbitrary constant quantity $\kappa$ here being inserted to make the argument of the logarithm dimensionless) 
are somewhat different from the expressions found in most thermodynamics texts. Thus we briefly review their derivation from standard thermodynamics and its adaptation to MHD (extending the discussion in \cite{Dewar_etal_08}).

First recall that, for a single-species ideal gas of absolute temperature $T_{\rm K}$ (in degrees Kelvin) whose atoms are of mass $m$ and number density is $n$, the mass density $\rho$ is $mn$ and the pressure $p$ is $nk_{\rm B}T_{\rm K}$, where $k_{\rm B}$ is Boltzmann's constant. The internal energy $U$ is $(3/2)nVk_{\rm B}T_{\rm K} = pV/(\gamma - 1)$, where $V$ is the volume of the system and $\gamma = 5/3$ is the ratio of specific heats. The statistical mechanical entropy $S_{\rm K}$ (in units such that a heat increment is $\d Q = T_{\rm K}\d S_{\rm K}$) is given by the Sackur--Tetrode equation
\begin{equation}\label{eq:SackTet}
	S_{\rm K} = Nk_{\rm B}\left\{\ln\left[\frac{V}{N}\left(\frac{4\pi m}{3h^2}\frac{U}{N}\right)^{3/2}\right] + \frac{5}{2}\right\}\;,
\end{equation}
where $N = nV$ is the number of particles and $h$ is Planck's constant.

In plasma physics, temperature $T$ is measured in energy units, i.e. $T = k_{\rm B}T_{\rm K}$, the corresponding entropy in energy units being $S_{\rm en} = S_{\rm K}/k_{\rm B}$ in order that $dQ = T\d S_{\rm en}$. Also there are two species, ions and electrons, to take into account, their number densities being denoted $n_{\rm i}$ and $n_{\rm e}$, respectively. If $Z_{\rm eff}$ is the effective ionization state then, to a very good approximation, $n_{\rm i}= n_{\rm e}/Z_{\rm eff}$ to maintain quasineutrality. Then the total pressure $p \equiv n_{\rm e}T_{\rm e} + n_{\rm i}T_{\rm i}$ becomes $n_{\rm e}(T_{\rm e} + T_{\rm i}/Z_{\rm eff})$. 

The MRxMHD assumption that current sheets on magnetic surfaces act as transport barriers is most justifiable if $T_{\rm e} \gg T_{\rm i}$, the small gyroradius of the electrons providing good confinement across magnetic field lines and their rapid motion along field lines providing fast thermal equilibration on magnetic surfaces and within the chaotic relaxation regions. Thus we henceforth assume the ions are cold, $T_{\rm i}/T_{\rm e} \ll 0$. However, the mass density is dominated by the ions because $m_{\rm e}/m_{\rm i} \ll 0$.

Simplifying notation by denoting $n_{\rm e}$ by $n$ and $T_{\rm e}$ by $T$, and defining an effective particle mass $m = m_{\rm i}/Z_{\rm eff}$, we summarize these approximations as
\begin{equation}\label{eq:denspress}
	\rho = nm \:\: \mathrm{and} \:\: \quad p = nT \;.
\end{equation}

To adapt standard thermodynamics we model the plasma as a monatomic gas at temperature $T$ made up of particles of mass $m$ [except in the de Broglie term in \eqref{eq:SackTet}, where we use $m_{\rm e}$] and write the Sackur--Tetrode equation \eqref{eq:SackTet} as
\begin{equation}\label{eq:SackTetEn}
	mS_{\rm en} = V s_m \;,
\end{equation}
where we have derived the entropy constraint density used in \eqref{eq:Li} as
\begin{eqnarray}\label{eq:sm}
	s_m &=&
	\frac{\rho}{\gamma - 1} \ln\left(\kappa\frac{p}{\rho^\gamma}\right)
	\;,
\end{eqnarray}
the hitherto arbitrary non-dimensionalizing constant $\kappa$ now being identified as
\begin{equation}\label{eq:kappadef}
	\kappa \equiv \frac{4\pi m_{\rm e}(me)^\gamma}{3(\gamma-1)h^2} \;.
\end{equation}

\section{Vector Magnetic Potential Boundary Constraints} 
\label{sec:Abdy}
In this appendix we seek to justify the holonomic constraint \eqref{eq:Abc} on a plasma or vacuum region boundary $\partial\Omega$. Also, to verify that magnetic fluxes are conserved under variation, we need to show line integrals $\oint\!\vrm{A}\dotv \d\vrm{l}$ around loops on the interface are invariant under displacements of the interface. While \eqref{eq:Abc} is as expected from the ideal MHD result, \cite{Bernstein_etal_58}, that $\delta\vrm{B} = \vrm{Q} \equiv \curl(\Delta\vrm{x}\cross\vrm{B})$, it needs to be justified for MRxMHD because we make no frozen-in-flux assumption other than the tangential-$\vrm{B}$ constraint. Within the subregions $\Omega_i$, the IMHD result $\delta\vrm{B} = \vrm{Q}$ does \emph{not} in general apply.

We first consider the problem of propagation of the tangential-$\vrm{B}$  condition \eqref{eq:tangential} on a time-dependent surface $\Gamma^t$ and then adapt the results to find the analogous $\epsilon$-variations at fixed $t$. 
First, from \eqref{eq:tangential} and \eqref{eq:tAdvSurf} we have (suppressing the superscripts $t$ unless needed to emphasize time dependence)
\begin{equation}\label{eq:AdvBdS}
	\frac{\d}{\d t}(\vrm{B}\dotv\d\vrm{S}) = \frac{\d\vrm{B}}{\d t}\dotv\d\vrm{S}
	- \vrm{B}\dotv(\grad\vrm{v})\dotv\d\vrm{S} = 0 \;.
\end{equation}
Dividing by $dS$ and using the definition $\d\vrm{B}/\d t \equiv \partial\vrm{B}/\partial t + \vrm{v}\dotv\grad\vrm{B}$ we thus find
\begin{equation}\label{eq:AdvBdS1}
	\vrm{n}\dotv\frac{\partial\vrm{B}}{\partial t}
	= \vrm{n}\dotv\curl(\vrm{v}\cross\vrm{B}) \:\:\mathrm{on}\:\:\Gamma\;.
\end{equation}
(NB This is obviously \emph{consistent} with the IMHD equation $\partial_t\vrm{B} = \curl(\vrm{v}\cross\vrm{B})$, but is derived completely generally and is thus applicable to MRxMHD as well.)

Substituting $\vrm{B} = \curl\vrm{A}$ in \eqref{eq:AdvBdS1} we easily find $\vrm{n}\dotv\curl(\partial_t\vrm{A} - \vrm{v}\cross\vrm{B}) = 0$, which is equivalent to
\begin{equation}\label{eq:AdvBdS1A}
	\frac{1}{|\grad\! f|}\divv\left[\grad\! f\cross\left(\frac{\partial\vrm{A}}{\partial t}
	- \vrm{v}\cross\vrm{B}\right)\right] = 0 \:\:\mathrm{on}\:\:\Gamma \;,
\end{equation}
where $f$ is a differentiable function such that $f(\vrm{x},t) = \const$ on $\Gamma^t$ (as in the last two paragraphs of Sec.~\ref{sec:Kinematics}) and we have used the identity $\curl\grad\! f \equiv 0$.

Although \eqref{eq:AdvBdS1A} is written using 3-dimensional Cartesian vector calculus notation, it applies only on the 2-dimensional surface $\Gamma$.  We now resolve this seeming paradox by transforming to a curvilinear coordinate system $f,g,h$ such that the basis vectors $\esup{f} \equiv \grad\! f = \vrm{n}|\grad\! f|$, $\esup{g} \equiv \grad g$, and $\esup{h} \equiv \grad h$ are linearly independent, so that the pair $h,g$ specifies a point on $\Gamma$: $f = \const$ and $\grad \equiv \esup{f}\partial_f + \esup{g}\partial_g + \esup{h}\partial_h$.

Using the identity $\mathcal{J}\, \divv\vrm{u} \equiv \partial_f(\mathcal{J}\esup{f}\dotv\vrm{u}) + \partial_g(\mathcal{J}\esup{g}\dotv\vrm{u}) + \partial_h(\mathcal{J}\esup{h}\dotv\vrm{u})$, where $\mathcal{J} \equiv 1/\esup{f}\dotv\esup{g}\cross\esup{h}$ and $\vrm{u}$ is an arbitrary vector field, and choosing $\vrm{u} = \grad\! f\cross(\partial_t\vrm{A} - \vrm{v}\cross\vrm{B})$, we find \eqref{eq:AdvBdS1A} is equivalent to
\begin{equation}\label{eq:AdvBdS2}
	\partial_g \left[\mathcal{J}\grad\! g\dotv\grad\! f\cross\left(\frac{\partial\vrm{A}}{\partial t} - \vrm{v}\cross\vrm{B}\right)\right]
     + \partial_h \left[\mathcal{J}\grad h\dotv\grad\! f\cross\left(\frac{\partial\vrm{A}}{\partial t} - \vrm{v}\cross\vrm{B}\right)\right]
     = 0 \;.
\end{equation} 
As the left-hand side does not contain the normal derivative $\partial_f$, it is a surface divergence operating purely on values of $\vrm{u}$ evaluated at the surface $\Gamma$. Thus the nature of the spatial dependence of $\vrm{B}(\vrm{x},t)$ off $\Gamma^t$ is immaterial to the evaluation of the boundary condition \eqref{eq:AdvBdS1A}---in particular $\vrm{B}$ does not need to have nested magnetic surfaces. [Likewise the off-surface dependence of $f(\vrm{x,t})$ is irrelevant, as $|\grad\! f|$ cancels in the product $\mathcal{J}\grad\! f = \vrm{n}/\vrm{n}\dotv\grad g\cross\grad h$.] We also note, using \eqref{eq:tangential}, that $\partial_t\vrm{A} - \vrm{v}\cross\vrm{B} = \grad\! f\cross[(\partial_t\vrm{A})_{\rm tgt}  + v_n\vrm{B}]$, so that only the tangential components of 
$\partial_t\vrm{A}$ and the normal velocity component $v _n \equiv \vrm{n}\dotv\vrm{v}$ contribute.

Clearly, the general solution of \eqref{eq:AdvBdS1A} is $\grad\! f\cross(\partial_t\vrm{A} - \vrm{v}\cross\vrm{B}) = \grad\! f\cross\grad\partial_t\chi$, where $\chi$ is an arbitrary gauge potential. Crossing both sides with $\vrm{n}/|\grad\! f|$ and rearranging gives the alternative form
\begin{equation}\label{eq:AdvBdS1AGS}
	\left(\frac{\partial\vrm{A}}{\partial t}\right)_{\rm tgt} 
	= \left(\vrm{v}\cross\vrm{B} + \grad\frac{\partial\chi}{\partial t}\right)_{\rm tgt} \;.
\end{equation}

Replacing $t$ with $\epsilon$ and $\vrm{v} \equiv \d\vrm{r}^t(\vrm{x})/\d t$ with $\d\vrm{r}^\epsilon(\vrm{x})/\d \epsilon$ in \eqref{eq:AdvBdS1AGS} and taking the limit as $\epsilon \to 0$ [cf. discussion after \eqref{eq:vtilde}] gives the desired variational holonomic constraint \eqref{eq:Abc}.

To show invariance of loop integrals $\oint\!\vrm{A}\dotv \d\vrm{l}$ under boundary and interface displacements, on surfaces that are not simply connected, we first show $\Delta(\vrm{A}\dotv \d\vrm{l}) = \Delta\vrm{A}\dotv \d\vrm{l} + \vrm{A}\dotv\Delta \d\vrm{l}$ is a complete differential on these surfaces. From the epsilon-flow analogue of \eqref{eq:tAdvLine}, $\Delta \d\vrm{l} = \d\vrm{l}\dotv\grad\Delta\vrm{x}$. Using the constraint \eqref{eq:Abc} we find
\begin{eqnarray}\label{eq:lineintvar}
	\Delta(\vrm{A}\dotv \d\vrm{l}) &=& \d\vrm{l}\dotv[\delta\vrm{A} + \Delta\vrm{x}\dotv\grad\vrm{A}
	+ (\grad\Delta\vrm{x})\dotv\vrm{A}] \nonumber\\
	&=&\d\vrm{l}\dotv[\Delta\vrm{x}\cross(\curl\vrm{A}) + \grad\delta\chi \nonumber\\
	&&\quad\quad + \Delta\vrm{x}\dotv\grad\vrm{A} + (\grad\Delta\vrm{x})\dotv\vrm{A}] \nonumber\\
	&=&  \d\vrm{l}\dotv\grad(\Delta\vrm{x}\dotv\vrm{A} + \grad\delta\chi) \;, 
\end{eqnarray}
which is a perfect differential as required. Thus, there is zero variation in line integrals around loops \emph{provided} we also require $\delta\chi$ to be single-valued.

\section{Acknowledgments}
One of the authors (RLD) gratefully acknowledges the support of The University of Tokyo and Princeton Plasma Physics Laboratory, during collaboration visits, and some travel support from Australian Research Council grant DP110102881. He also acknowledges useful discussions with Philip Morrison.  The work of ZY was supported under JSPS grant KAKENHI 23224014 and that of AB and SRH was supported under US DOE grant DE-AC02-09CH11466. The plots were made using Mathematica 10, \cite{Mathematica10}.

\bibliographystyle{jpp}
\bibliography{RLDBibDeskPapers}

\begin{thebibliography}{66}
\expandafter\ifx\csname natexlab\endcsname\relax\def\natexlab#1{#1}\fi

\bibitem[Araki(2015)]{Araki_15}
{\sc Araki, K.} 2015 Differential-geometrical approach to the dynamics of
  dissipationless incompressible {H}all magnetohydrodynamics: I. {L}agrangian
  mechanics on semidirect product of two volume preserving diffeomorphisms and
  conservation laws. {\em Journal of Physics A: Math. Theoretical\/} {\bf 48},
  175501--1--16.

\bibitem[Arnold \& Khesin(1998)]{Arnold_Khesin_98}
{\sc Arnold, V.~I. \& Khesin, B.~A.} 1998 {\em Topological Methods in
  Hydrodynamics\/}, {\em Applied Mathematical Sciences\/}, vol. 125. New York:
  Springer.

\bibitem[Berger(1999)]{Berger_99}
{\sc Berger, M.~A.} 1999 Introduction to magnetic helicity. {\em Plasma Phys.
  Control. Fusion\/} {\bf 41}, B167--B175.

\bibitem[Bernstein {\em et~al.\/}(1958)Bernstein, Frieman, Kruskal \&
  Kulsrud]{Bernstein_etal_58}
{\sc Bernstein, I.~B., Frieman, E.~A., Kruskal, M.~D. \& Kulsrud, R.~M.} 1958
  An energy principle for hydromagnetic stability problems. {\em Proc. Roy.
  Soc. London Ser. A\/} {\bf 244}, 17--40.

\bibitem[Bevir \& Gray(1982)]{Bevir_Gray_82}
{\sc Bevir, M.~K. \& Gray, J.~W.} 1982 Relaxation, flux conservation and quasi
  steady state pinches. In {\em Proceedings of the Reversed Field Pinch Theory
  Workshop, Los Alamos, NM, USA, 28 Apr - 2 May 1980\/} (ed. H.~R. Lewis), pp.
  176--180. Los Alamos National Laboratory, Los Alamos National Laboratory.

\bibitem[Bhattacharjee \& Dewar(1982)]{Bhattacharjee_Dewar_82}
{\sc Bhattacharjee, A. \& Dewar, R.~L.} 1982 Energy principle with global
  invariants. {\em Phys. Fluids\/} {\bf 25}, 887--897.

\bibitem[Bhattacharjee {\em et~al.\/}(1995)Bhattacharjee, Hayashi, Hegna,
  Nakajima \& Sato]{Bhattacharjee_etal_95}
{\sc Bhattacharjee, A., Hayashi, T., Hegna, C.~C., Nakajima, N. \& Sato, T.}
  1995 Theory of {pressure-induced} islands and {self-healing} in
  {three-dimensional} toroidal magnetohydrodynamic equilibria. {\em Phys.
  Plasmas\/} {\bf 2}, 883--888.

\bibitem[Boozer \& Pomphrey(2010)]{Boozer_Pomphrey_10}
{\sc Boozer, A.~H. \& Pomphrey, N.} 2010 Current density and plasma
  displacement near perturbed rational surfaces. {\em Phys. Plasmas\/} {\bf
  17}, 110707--1--4.

\bibitem[Cary \& Kotschenreuther(1985)]{Cary_Kotschenreuther_85}
{\sc Cary, J.~R. \& Kotschenreuther, M.} 1985 Pressure induced islands in
  {three-dimensional} toroidal plasma. {\em Phys. Fluids\/} {\bf 28},
  1392--1401.

\bibitem[Comisso {\em et~al.\/}(2015{\natexlab{{\em a\/}}})Comisso, Grasso \&
  Waelbroeck]{Comisso_Grasso_Waelbroeck_15a}
{\sc Comisso, L., Grasso, D. \& Waelbroeck, F.~L.} 2015{\natexlab{{\em a\/}}}
  Extended theory of the {T}aylor problem in the plasmoid-unstable regime. {\em
  Phys. Plasmas\/} {\bf 22}, 042109--1--12.

\bibitem[Comisso {\em et~al.\/}(2015{\natexlab{{\em b\/}}})Comisso, Grasso \&
  Waelbroeck]{Comisso_Grasso_Waelbroeck_15b}
{\sc Comisso, L., Grasso, D. \& Waelbroeck, F.~L.} 2015{\natexlab{{\em b\/}}}
  Phase diagrams of forced magnetic reconnection in {T}aylor's model. {\em
  Submitted for publication in J. Plasma Phys.\/} p. 13 pages, 2015 Workshop
  ``Complex plasma phenomena in the laboratory and in the universe''.

\bibitem[Cordoba \& Marliani(2000)]{Cordoba_Marliani_00}
{\sc Cordoba, D. \& Marliani, C.} 2000 Evolution of current sheets and
  regularity of ideal incompressible magnetic fluids in 2d. {\em Communications
  on Pure and Applied Mathematics, Vol.\/} {\bf LIII}, 0512---0524.

\bibitem[Dennis {\em et~al.\/}(2013{\natexlab{{\em a\/}}})Dennis, Hudson, Dewar
  \& Hole]{Dennis_Hudson_Dewar_Hole_13}
{\sc Dennis, G.~R., Hudson, S.~R., Dewar, R.~L. \& Hole, M.~J.}
  2013{\natexlab{{\em a\/}}} The infinite interface limit of multiple-region
  relaxed mhd. {\em Phys. Plasmas\/} {\bf 20}, 032509--1--6.

\bibitem[Dennis {\em et~al.\/}(2014)Dennis, Hudson, Dewar \&
  Hole]{Dennis_Hudson_Dewar_Hole_14a}
{\sc Dennis, G.~R., Hudson, S.~R., Dewar, R.~L. \& Hole, M.~J.} 2014
  Multi-region relaxed magnetohydrodynamics with flow. {\em Phys. Plasmas\/}
  {\bf 21}, 042501--1--9.

\bibitem[Dennis {\em et~al.\/}(2013{\natexlab{{\em b\/}}})Dennis, Hudson,
  Terranova, Franz, Dewar \& Hole]{Dennis_etal_13}
{\sc Dennis, G.~R., Hudson, S.~R., Terranova, D., Franz, P., Dewar, R.~L. \&
  Hole, M.~J.} 2013{\natexlab{{\em b\/}}} A minimally constrained model of
  self-organized helical states in reversed-field pinches. {\em Phys. Rev.
  Lett.\/} {\bf 111}, 055003--1--5.

\bibitem[Dewar(1970)]{Dewar_70}
{\sc Dewar, R.~L.} 1970 Interaction between hydromagnetic waves and a
  time-dependent, inhomogeneous medium. {\em Phys. Fluids\/} {\bf 13},
  2710--2720.

\bibitem[Dewar(1976)]{Dewar_76b}
{\sc Dewar, R.~L.} 1976 Renormalised canonical perturbation theory for
  stochastic propagators. {\em J. Phys. A: Math. Gen.\/} {\bf 9}, 2043--2057.

\bibitem[Dewar(1978)]{Dewar_78a}
{\sc Dewar, R.~L.} 1978 Hamilton's principle for a hydromagnetic fluid with a
  free boundary. {\em Nucl. Fusion\/} {\bf 18}, 1541--1553.

\bibitem[Dewar {\em et~al.\/}(2013)Dewar, Bhattacharjee, Kulsrud \&
  Wright]{Dewar_Bhattacharjee_Kulsrud_Wright_13}
{\sc Dewar, R.~L., Bhattacharjee, A., Kulsrud, R.~M. \& Wright, A.~M.} 2013
  Plasmoid solutions of the {H}ahm--{K}ulsrud--{T}aylor equilibrium model. {\em
  Phys. Plasmas\/} {\bf 20}, 082103--1--7.

\bibitem[Dewar {\em et~al.\/}(2008)Dewar, Hole, McGann, Mills \&
  Hudson]{Dewar_etal_08}
{\sc Dewar, R.~L., Hole, M.~J., McGann, M., Mills, R. \& Hudson, S.~R.} 2008
  Relaxed plasma equilibria and entropy-related plasma self-organization
  principles. {\em Entropy\/} {\bf 10}, 621--634.

\bibitem[Dombre {\em et~al.\/}(1986)Dombre, Frisch, Greene, H{\'e}non, Mehr \&
  Soward]{Dombre_etal_1986}
{\sc Dombre, T., Frisch, U., Greene, J.~M., H{\'e}non, M., Mehr, A. \& Soward,
  A.~M.} 1986 Chaotic streamlines in the abc flows. {\em J. Fluid Mech.\/} {\bf
  167}, 353--391.

\bibitem[Freidberg(1982)]{Freidberg_82}
{\sc Freidberg, J.~P.} 1982 Ideal magnetohydrodynamic theory of magnetic fusion
  systems. {\em Rev. Mod. Phys.\/} {\bf 54}, 801--902.

\bibitem[Freidberg(1987)]{Freidberg_87}
{\sc Freidberg, J.~P.} 1987 {\em Ideal Magnetohydrodynamics\/}. New York:
  Plenum Press.

\bibitem[Frieman \& Rotenberg(1960)]{Frieman_Rotenberg_60}
{\sc Frieman, E. \& Rotenberg, M.} 1960 On hydromagnetic stability of
  stationary equilibria. {\em Rev. Mod. Phys.\/} {\bf 32}, 898--902.

\bibitem[Goldstein(1980)]{Goldstein_80}
{\sc Goldstein, H.} 1980 {\em Classical {M}echanics\/}, 2nd edn. Reading,
  Mass., USA: Addison-Wesley.

\bibitem[Grad(1967)]{Grad_67}
{\sc Grad, H.} 1967 Toroidal containment of a plasma. {\em Phys. Fluids\/} {\bf
  10}, 137--154.

\bibitem[Hahm \& Kulsrud(1985)]{Hahm_Kulsrud_85}
{\sc Hahm, T.~S. \& Kulsrud, R.~M.} 1985 Forced magnetic reconnection. {\em
  Phys. Fluids\/} {\bf 28}, 2412--2418.

\bibitem[Hameiri(2014)]{Hameiri_14}
{\sc Hameiri, E.} 2014 Some improvements in the theory of plasma relaxation.
  {\em Phys. Plasmas\/} {\bf 21}, 044503--1--5.

\bibitem[Hegna \& Bhattacharjee(1989)]{Hegna_Bhattacharjee_89}
{\sc Hegna, C.~C. \& Bhattacharjee, A.} 1989 Magnetic island formation in
  three‐dimensional plasma equilibria. {\em Phys. Fluids B\/} {\bf 1},
  392--397.

\bibitem[Helander(2014)]{Helander_14}
{\sc Helander, P.} 2014 Theory of plasma confinement in non-axisymmetric
  magnetic fields. {\em Rep. Prog. Phys.\/} {\bf 77}, 087001--1--35.

\bibitem[Hole {\em et~al.\/}(2007)Hole, Hudson \& Dewar]{Hole_Hudson_Dewar_07}
{\sc Hole, M.~J., Hudson, S.~R. \& Dewar, R.~L.} 2007 Equilibria and stability
  in partially relaxed plasma--vacuum systems. {\em Nucl. Fusion\/} {\bf 47},
  746--753.

\bibitem[Hosking \& Dewar(2015)]{Hosking_Dewar_15}
{\sc Hosking, R.~J. \& Dewar, R.~L.} 2015 {\em Fundamental Fluid Mechanics and
  Magnetohydrodynamics\/}. Singapore: Springer Singapore.

\bibitem[Hudson {\em et~al.\/}(2012)Hudson, Dewar, Dennis, Hole, McGann, von
  Nessi \& Lazerson]{Hudson_etal_12b}
{\sc Hudson, S.~R., Dewar, R.~L., Dennis, G., Hole, M.~J., McGann, M., von
  Nessi, G. \& Lazerson, S.} 2012 Computation of multi-region relaxed
  magnetohydrodynamic equilibria. {\em Phys. Plasmas\/} {\bf 19},
  112502--1--18.

\bibitem[Hudson {\em et~al.\/}(2007)Hudson, Hole \&
  Dewar]{Hudson_Hole_Dewar_07}
{\sc Hudson, S.~R., Hole, M.~J. \& Dewar, R.~L.} 2007 Eigenvalue problems for
  {B}eltrami fields arising in a three-dimensional toroidal magnetohydrodynamic
  equilibrium problem. {\em Phys. Plasmas\/} {\bf 14}, 052505--1--12.

\bibitem[Jensen \& Chu(1984)]{Jensen_Chu_84}
{\sc Jensen, T.~H. \& Chu, M.~S.} 1984 Current drive and helicity injection.
  {\em Phys. Fluids\/} {\bf 27}, 2881--2885.

\bibitem[K.~Charidakos {\em et~al.\/}(2014)K.~Charidakos, Lingam, Morrison,
  White \& Wurm]{Charidakos_etal_14}
{\sc K.~Charidakos, I., Lingam, M., Morrison, P.~J., White, R.~L. \& Wurm, A.}
  2014 Action principles for extended magnetohydrodynamic models. {\em Phys.
  Plasmas\/} {\bf 21}, 092118--1--12.

\bibitem[Kruskal \& Kulsrud(1958)]{Kruskal_Kulsrud_58}
{\sc Kruskal, M.~D. \& Kulsrud, R.~M.} 1958 Equilibrium of a magnetically
  confined plasma in a toroid. {\em Phys. Fluids\/} {\bf 1}, 265--274.

\bibitem[Loizu {\em et~al.\/}(2015{\natexlab{{\em a\/}}})Loizu, Hudson,
  Bhattacharjee \& Helander]{Loizu_etal_15a}
{\sc Loizu, J., Hudson, S., Bhattacharjee, A. \& Helander, P.}
  2015{\natexlab{{\em a\/}}} Magnetic islands and singular currents at rational
  surfaces in three-dimensional magnetohydrodynamic equilibria. {\em Phys.
  Plasmas\/} {\bf 22}, 022501--1--12.

\bibitem[Loizu {\em et~al.\/}(2015{\natexlab{{\em b\/}}})Loizu, Hudson,
  Bhattacharjee, Lazerson \& Helander]{Loizu_etal_15b}
{\sc Loizu, J., Hudson, S.~R., Bhattacharjee, A., Lazerson, S. \& Helander, P.}
  2015{\natexlab{{\em b\/}}} Existence of three-dimensional ideal-{MHD}
  equilibria with current sheets. {\em Phys. Plasmas\/} {\bf 22}, 090704--1--5.

\bibitem[Longcope \& Strauss(1993)]{Longcope_Strauss_93}
{\sc Longcope, D.~W. \& Strauss, H.~R.} 1993 The coalescence instability and
  the development of current sheets in two-dimensional magnetohydrodynamics.
  {\em Phys. Fluids B\/} {\bf 5}, 2858--2869.

\bibitem[McGann(2013)]{McGann_13}
{\sc McGann, M.} 2013 Hamilton-{J}acobi theory for connecting equilibrium
  magnetic fields across a toroidal surface supporting a plasma pressure
  discontinuity. Ph.d. thesis, Australian National University, Canberra ACT
  0200, Australia Australia.

\bibitem[McGann {\em et~al.\/}(2010)McGann, Hudson, Dewar \& {von
  Nessi}]{McGann_Hudson_Dewar_vonNessi_10}
{\sc McGann, M., Hudson, S.~R., Dewar, R.~L. \& {von Nessi}, G.} 2010
  Hamilton--{J}acobi theory for continuation of magnetic field across a
  toroidal surface supporting a plasma pressure discontinuity. {\em Phys.
  Letts. A\/} {\bf 374}, 3308--3314.

\bibitem[Mills {\em et~al.\/}(2009)Mills, Hole \& Dewar]{Mills_Hole_Dewar_09}
{\sc Mills, R., Hole, M.~J. \& Dewar, R.~L.} 2009 Magnetohydrodynamic stability
  of plasmas with ideal and relaxed regions. {\em J. Plasma Phys.\/} {\bf 75},
  637--659.

\bibitem[Morrison(1998)]{Morrison_98}
{\sc Morrison, P.~J.} 1998 Hamiltonian description of the ideal fluid. {\em
  Rev. Mod. Phys.\/} {\bf 70}, 467--521.

\bibitem[Newcomb(1962)]{Newcomb_62}
{\sc Newcomb, W.~A.} 1962 Lagrangian and {H}amiltonian methods in
  magnetohydrodynamics. {\em Nucl. Fusion Suppl.\/} {\bf Part 2}, 451--463.

\bibitem[Padhye \& Morrison(1996{\natexlab{{\em a\/}}})]{Padhye_Morrison_96a}
{\sc Padhye, N. \& Morrison, P.~J.} 1996{\natexlab{{\em a\/}}} Fluid element
  relabeling symmetry. {\em Phys. Lett. A\/} {\bf 219}, 287--292.

\bibitem[Padhye \& Morrison(1996{\natexlab{{\em b\/}}})]{Padhye_Morrison_96b}
{\sc Padhye, N. \& Morrison, P.~J.} 1996{\natexlab{{\em b\/}}} Relabeling
  symmetries in hydrodynamics and magnetohydrodynamics. {\em Plasma Phys.
  Reports\/} {\bf 22}, 869--877.

\bibitem[Parker(1994)]{Parker_94}
{\sc Parker, E.~N.} 1994 {\em Spontaneous Current Sheets in Magnetic Fields
  with Applications to Stellar X-Rays\/}. {\em International Series in
  Astronomy and Astrophysics\/} 1. New York: Oxford University Press.

\bibitem[{Potter}(1976)]{Potter_76}
{\sc {Potter}, D.} 1976 {\em {Waterbag methods in magnetohydrodynamics}\/},
  {\em Methods in Computational Physics\/}, vol.~16, pp. 43--83. New York:
  Academic Press.

\bibitem[Qin {\em et~al.\/}(2012)Qin, Liu, Li \& Squire]{Qin_Liu_Li_Squire_12}
{\sc Qin, H., Liu, W., Li, H. \& Squire, J.} 2012 Woltjer-{T}aylor state
  without {T}aylor's {C}onjecture: Plasma relaxation at all wavelengths. {\em
  Phys. Rev. Lett.\/} {\bf 109}, 235001--1--5.

\bibitem[Rusbridge(1991)]{Rusbridge_91}
{\sc Rusbridge, M~G} 1991 The relationship between the `tangled discharge' and
  `dynamo' models of the magnetic relaxation process. {\em Plasma Physics and
  Controlled Fusion\/} {\bf 33}, 1381--1389.

\bibitem[Salmon(1988)]{Salmon_88}
{\sc Salmon, R.} 1988 Hamiltonian fluid mechanics. {\em Ann. Rev. Fluid
  Mech.\/} {\bf 20}, 225--256.

\bibitem[Smiet {\em et~al.\/}(2015)Smiet, Candelaresi, Thompson, Swearngin,
  Dalhuizen \& Bouwmeester]{Smiet_etal_15}
{\sc Smiet, C.~B., Candelaresi, S., Thompson, A., Swearngin, J., Dalhuizen,
  J.~W. \& Bouwmeester, D.} 2015 Self-organizing knotted magnetic structures in
  plasma. {\em Phys. Rev. Letters\/} {\bf 115}, 095001--1--5.

\bibitem[Stott {\em et~al.\/}(1977)Stott, Wilson \&
  Gibson]{Stott_Wilson_Gibson_77}
{\sc Stott, P.~E., Wilson, C.~M. \& Gibson, A.} 1977 The bundle divertor --
  part {I}: Magnetic configuration. {\em Nucl. Fusion\/} {\bf 17}, 481--496.

\bibitem[Taylor(1974)]{Taylor_74}
{\sc Taylor, J.~B.} 1974 Relaxation of toroidal plasma and generation of
  reverse magnetic fields. {\em Phys. Rev. Lett.\/} {\bf 33}, 1139--1141.

\bibitem[Taylor(1986)]{Taylor_86}
{\sc Taylor, J.~B.} 1986 Relaxation and magnetic reconnection in plasmas. {\em
  Rev. Mod. Phys.\/} {\bf 58}, 741--763.

\bibitem[Waelbroeck(1989)]{Waelbroeck_89}
{\sc Waelbroeck, F.~L.} 1989 Current sheets and nonlinear growth of the $m=1$
  kink-tearing mode. {\em Phys. Plasmas B\/} {\bf 1}, 2372--2380.

\bibitem[Wang \& Bhattacharjee(1995)]{Wang_Bhattacharjee_95}
{\sc Wang, X. \& Bhattacharjee, A.} 1995 Nonlinear dynamics of the $m=1$
  kink-tearing instability in a modified magnetohydrodynamic model. {\em Phys.
  Plasmas\/} {\bf 2}, 171--181.

\bibitem[Webb {\em et~al.\/}(2014{\natexlab{{\em a\/}}})Webb, Dasgupta1,
  McKenzie, Hu \& Zank]{Webb_etal_2014_I}
{\sc Webb, G.~M., Dasgupta1, B., McKenzie, J.~F., Hu, Q. \& Zank, G.~P.}
  2014{\natexlab{{\em a\/}}} Local and nonlocal advected invariants and
  helicities in magnetohydrodynamics and gas dynamics i: {L}ie dragging
  approach. {\em J. Phys. A., Math. and Theor.\/} {\bf 47}, 095501--1--33.

\bibitem[Webb {\em et~al.\/}(2014{\natexlab{{\em b\/}}})Webb, Dasgupta1,
  McKenzie, Hu \& Zank]{Webb_etal_2014_II}
{\sc Webb, G.~M., Dasgupta1, B., McKenzie, J.~F., Hu, Q. \& Zank, G.~P.}
  2014{\natexlab{{\em b\/}}} Local and nonlocal advected invariants and
  helicities in magnetohydrodynamics and gas dynamics ii: {N}oether's theorems
  and {C}asimirs. {\em J. Phys. A., Math. and Theor.\/} {\bf 47},
  095502--1--31.

\bibitem[Webb \& Zank(2007)]{Webb_Zank_07}
{\sc Webb, G.~M. \& Zank, G.~P.} 2007 Fluid relabelling symmetries, {L}ie point
  symmetries and the {L}agrangian map in magnetohydrodynamics and gas dynamics.
  {\em J. Phys. A: Math. Theor.\/} {\bf 40}, 545--579.

\bibitem[White(2013)]{White_13}
{\sc White, R.~B.} 2013 Representation of ideal magnetohydrodynamic modes. {\em
  Phys. Plasmas\/} {\bf 20}, 022105--1--4.

\bibitem[{Wolfram Research, Inc.}(2015)]{Mathematica10}
{\sc {Wolfram Research, Inc.}} 2015 {\em Mathematica, Version 10.1\/}.
  Champaign, Illinois, USA: Wolfram Research.

\bibitem[Woltjer(1958)]{Woltjer_58a}
{\sc Woltjer, L.} 1958 A theorem on force-free magnetic fields. {\em Proc. Nat.
  Acad. Sci. (U.S.)\/} {\bf 44}, 489--491.

\bibitem[Yoshida \& Dewar(2012)]{Yoshida_Dewar_12}
{\sc Yoshida, Z. \& Dewar, R.~L.} 2012 Helical bifurcation and tearing mode in
  a plasma --- a description based on {C}asimir foliation. {\em J. Phys. A:
  Math. Gen.\/} {\bf 45}, 365502--1--36.

\bibitem[Yoshida \& Giga(1990)]{Yoshida_Giga_90}
{\sc Yoshida, Z. \& Giga, Y.} 1990 Remarks on spectra of operator rot. {\em
  Math. Z.\/} {\bf 204}, 235--245.

\end{thebibliography}

\end{document}